\documentclass[iop,apj,numberappendix,appendixfloats,twocolappendix]{emulateapj}
\usepackage{graphicx}
\LongTables
\def \cmm  {cm$^{-2}$}

\def \lya  {Ly$\alpha$}

\def \ly5  {Ly-5}
\def \ly6  {Ly-6}
\def \ly7  {Ly-7}

\def \mlnovi  {\log N_{\rm OVI}}

\newcommand{\km}{${\rm km\,s}^{-1}$}

\newcommand{\hst}{{\em HST}}
\newcommand{\fuse}{{\em FUSE}}

\def\lesssim{\mathrel{\hbox{\rlap{\hbox{%
 \lower4pt\hbox{$\sim$}}}\hbox{$<$}}}}
\def\gtrsim{\mathrel{\hbox{\rlap{\hbox{%
 \lower4pt\hbox{$\sim$}}}\hbox{$>$}}}}
\let\la=\lesssim                
\let\ga=\gtrsim

\newcommand{\hi}{H$\;${\small\rm I}\relax}
\newcommand{\hii}{H$\;${\small\rm II}\relax}

\newcommand{\cii}{C$\;${\small\rm II}\relax}
\newcommand{\ciii}{C$\;${\small\rm III}\relax}
\newcommand{\civ}{C$\;${\small\rm IV}\relax}

\newcommand{\oi}{O$\;${\small\rm I}\relax}

\newcommand{\ovi}{O$\;${\small\rm VI}\relax}

\newcommand{\sii}{S$\;${\small\rm II}\relax}
\newcommand{\siii}{Si$\;${\small\rm II}\relax}
\newcommand{\siiii}{Si$\;${\small\rm III}\relax}

\newcommand{\siiv}{Si$\;${\small\rm IV}\relax}

\newcommand{\hit}{H$\;${\scriptsize\rm I}\relax}

\newcommand{\ciit}{C$\;${\scriptsize\rm II}\relax}

\newcommand{\ciiit}{C$\;${\scriptsize\rm III}\relax}
\newcommand{\civt}{C$\;${\scriptsize\rm IV}\relax}

\newcommand{\oit}{O$\;${\scriptsize\rm I}\relax}

\newcommand{\ovit}{O$\;${\scriptsize\rm VI}\relax}

\newcommand{\mgiit}{Mg$\;${\scriptsize\rm II}\relax}

\newcommand{\siiit}{Si$\;${\scriptsize\rm II}\relax}

\newcommand{\siiiit}{Si$\;${\scriptsize\rm III}\relax}

\newcommand{\siivt}{Si$\;${\scriptsize\rm IV}\relax}

\submitted{Accepted by the ApJ}
\shortauthors{Lehner, Howk, \& Wakker}
\shorttitle{The Circumgalactic Medium of M31}

\begin{document}

\title{Evidence for a Massive, Extended Circumgalactic Medium Around the Andromeda Galaxy\altaffilmark{1}}

\author{
Nicolas Lehner\altaffilmark{2},
J. Christopher Howk\altaffilmark{2}, and
Bart P. Wakker\altaffilmark{3} 
}

\altaffiltext{1}{Based on observations made with the NASA/ESA Hubble Space Telescope,
obtained at the Space Telescope Science Institute, which is operated by the
Association of Universities for Research in Astronomy, Inc. under NASA
contract No. NAS5-26555.}
\altaffiltext{2}{Department of Physics, University of Notre Dame, 225 Nieuwland Science Hall, Notre Dame, IN 46556}
\altaffiltext{3}{Supported by NASA/NSF, affiliated with the Department of Astronomy, University of Wisconsin, 475 N. Charter Street, Madison, WI 53706}

\begin{abstract}
We demonstrate the presence of an extended and massive circumgalactic medium (CGM) around Messier 31 using archival \hst\ COS ultraviolet spectroscopy of 18 QSOs projected within two virial radii of M31 ($R_{\rm vir} =300$ kpc). We detect absorption from \siiii\ at $-300 \la v_{\rm LSR} \la -150$ \km\ toward all 3 sightlines at $R \la 0.2 R_{\rm vir}$,   3 of 4 sightlines at  $0.8\la R/R_{\rm vir}\la 1.1$, and possibly 1 of 11 at  $1.1< R/R_{\rm vir}\la 1.8$. We present several arguments that the gas at these velocities observed in these directions originates from the CGM of M31 rather than the Local Group or Milky Way CGM or Magellanic Stream.  We show that the dwarf galaxies located in the CGM of M31 have very similar velocities over similar projected distances from M31. We find a non-trivial relationship only at these velocities between the column densities ($N$) of all the ions and $R$, whereby $N$ decreases with increasing $R$.  Singly ionized species are only detected in the inner CGM of M31 at $<0.2 R_{\rm vir}$. At $R<0.8 R_{\rm vir}$, the covering fraction is close to unity for \siiii\ and \civ\ ($f_{\rm c}\sim 60\%$--$97\%$ at the 90\% confidence level), but drops to $f_{\rm c}\la 10$--$20\%$ at $R\ga R_{\rm vir}$. We show that the M31 CGM gas is bound,  multiphase, predominantly ionized (i.e., \hii\,$\gg$\,\hi), and becomes more highly ionized gas at larger $R$. We estimate using \siii, \siiii, and \siiv\ a CGM metal mass of  at least $2\times 10^6$ M$_\sun$ and gas mass of $\ga 3 \times 10^9\,(Z_\sun/Z)$ M$_\sun$ within $0.2 R_{\rm vir}$, and possibly a factor $\sim$10 larger within $R_{\rm vir}$, implying substantial metal and gas masses in the CGM of M31. Compared with galaxies in the COS-Halos survey, the CGM of M31 appears to be quite typical for a $L^*$ galaxy. 
\end{abstract}


\section{Introduction}\label{s-intro}
The circumgalactic medium (CGM), loosely defined as the diffuse gas between the thick disk of galaxies and about a virial radius of galaxies is the scene where large-scale inflow and outflow from galaxies takes place. The competition between these processes is thought to shape galaxies and drive their evolution  \citep[e.g.,][]{keres05,dekel06,cafg11,putman12}. Observations of the properties of the CGM are therefore critical to test theories of galaxy evolution. Recent discoveries using high quality ultraviolet observations show that the CGM is a pivotal component of galaxies with significant mass of baryons and metals \citep[e.g.,][]{wakker09,chen10,prochaska11,tumlinson11,churchill12,kacprzak12, kacprzak14,werk13,werk14,lehner13,stocke13,peeples14,werk14,liang14,bordoloi14}. 

Observations of the CGM at $z>0$ typically provide only average properties of the CGM along one sightline \citep[in some rare cases, 2--3 sightlines, e.g.,][]{keeney13}, and therefore the CGM properties such as gas kinematics, metal mass distribution, and ionization states as a function of galaxy geometry and properties are not well constrained. One way to alleviate in part this issue is to determine the properties of the CGM for  similar type or mass of galaxies with sightlines piercing their CGM at various impact parameters. This is the strategy used in the COS-Halos survey \citep{tumlinson13}.   With a controlled sample of $L^*$ galaxies, the COS-Halos survey  provided strong evidence for extended highly ionized CGM by demonstrating that typical  star-forming $L^*$ galaxies have \ovi\ column densities $N_{\rm OVI} \ga 10^{14.3}$ \cmm, while more passive $L^*$ galaxies show weaker or no \ovi\ absorption in their CGM \citep{tumlinson11}. While this experiment has been extremely fruitful \citep{thom11,thom12,tumlinson11,tumlinson13,werk13,werk14,peeples14}, the COS-Halos galaxies have enough spread in their properties that even collectively they do not mimic a single galaxy. 

With several tens of QSO sightlines going through the Milky Way (MW) CGM \citep[e.g.,][]{savage03,sembach03,wakker03,wakker12,fox06,shull09,lehner12}, the MW would appear a perfect candidate for a ``zoom-in'' experiment, i.e., in which we can study the CGM along different sightlines of a single galaxy. However, our understanding of the MW CGM has remained somewhat limited by our position within the MW disk. The high-velocity clouds (HVCs, clouds that have typically $|v_{\rm LSR}|\ge 90$ \km\ at $|b|\ga 20\degr$, e.g., \citealt{wakker01} ) that cover the Galactic sky were thought to possibly probe the extended MW CGM. Their distances are now largely determined (\citealt{ryans97,wakker01,thom06,thom08,wakker07,wakker08,lehner10,smoker11} for \hi\ HVC complexes and \citealt{lehner11}; \citealt{lehner12}  for the diffuse ionized HVCs, iHVCs), but this  created another puzzle since they  place most of the \hi\ HVCs and iHVCs within 5--15 kpc of the sun, not at 100--300 kpc,  the expected size of the MW CGM, and hence the HVCs only represent a comparatively very small mass (since $M\propto d^2$). Only the Magellanic Stream \citep[MS, e.g.,][]{putman98,bruns05,nidever08,fox14} is more distant, possibly extending to 80--200 kpc \citep{besla12}, providing some indirect evidence for an extended corona around the MW \citep[e.g.,][]{stan02,sembach03,fox13,fox14}. 

The high \ovi\ column density found by COS-Halos is another element that shows that the iHVCs do not probe the extended MW CGM since $N_{\rm OVI}$ for the MW iHVCs is on average a factor 5 times smaller \citep[see the results in][]{sembach03,fox06}. Only if the entire MW thick disk and halo absorption is integrated, would $N_{\rm OVI}$ in the MW approach $10^{14.3}$ \cmm\ (i.e., by combining the results of \citealt{savage03} and \citealt{sembach03}). This would mean that  most of the column density of the MW CGM might be hidden in part in the low-velocity gas often associated with the thin and thick disks (see  \citealt{peek09} and Y. Zheng et al. 2015, in preparation). However, it is also plausible that the MW CGM has properties that are different from $z\sim 0.5$ $L^*$ galaxies. 

Studies of the gas content of nearby galaxies offer major advantages over both the MW and higher redshift galaxies. Nearby galaxies span a large angular extent and can be studied over multiple lines-of-sight and offer a direct mapping between the stellar distribution and the gas content. This experiment has been conducted for the Large Magellanic Cloud (\citealt{howk02}; \citealt{lehner07}; \citealt{lehner09}; \citealt{barger14}; N. Lehner et al., 2015 in prep.), showing in particular the presence of large-scale outflows from a sub-$L^*$ galaxy feeding in metals the CGM of a $L*$ galaxy (the MW). 

The $L^*$ galaxy that can be observed with the most detail is the Andromeda Nebula (M31). The stellar disk and halo of M31 have been subject to intense study \citep[e.g.,][]{brown06,mcconnachie09,gilbert12,gilbert14,dalcanton12}, with well-determined local and global properties, including its inclination \citep{walterbos87}, stellar and virial masses \citep{geehan06,marel12}, dust and ISM disk mass \citep{draine14}, rotation curve \citep[e.g.,][]{chemin09}, and its galaxy satellites \citep{mcconnachie12}. These studies all imply that M31 is fairly typical of massive star-forming galaxies,which has undergone several major interactions with its satellites \citep[e.g.,][]{gordon06,putman09}, and possibly in a phase of transformation into a red galaxy \citep{mutch11,davidge12}. The specific star-formation rate of M31, sSFR\,$\equiv {\rm SFR/M_\star} = (5\pm 1) \times 10^{-12}$ yr$^{-1}$ \citep[using the stellar mass M$_\star$ and SFR from][]{geehan06,kang09}, places M31 just between the passively evolving and star forming galaxies in COS-Halos \citep{tumlinson11}. As we show in this paper, the value of $N_{\rm OVI}$ in the CGM of M31 and the sSFR of M31 are also consistent with M31 being in the ``green valley''. 

Parallel to this intensive observational effort, there is also a major theoretical effort to understand the two massive galaxies, the MW and M31, of the Local Group (LG) \citep{richter12,garrison14,nuza14}. With this large amount of empirical results, M31 could become a benchmark for assessing the validity of the physics implemented in these simulations. This requires to have also knowledge of its extended diffuse ionized CGM. Both deep and shallower observations of the \hi\ 21-cm emission have reported detections of \hi\ clouds mostly within 50 kpc \citep{thilker04,westmeier05,westmeier08}, and at farther distances along the M31--M33 axis \citep{braun04}. The \hi\ 21-cm emission $49\arcmin$ resolution map in \citet{braun04} gave the impression of an \hi\ bridge between M31 and M33, but subsequent deep Green Bank Telescope (GBT) $9\farcm1$ resolution observations show that they appear to be small concentrations of \hi\ with $\sim 10^5$ M$_\sun$ on scale of a few kpc \citep{lockman12,wolfe13}. As for the MW, these \hi\ clouds might only be the tip of the iceberg and \hi\ alone does not provide information on the gas-phases, the metal content, and hence the total metal and baryon masses of the gas in the CGM of M31. 

We therefore mined the \hst\ Cosmic Origins Spectrograph (COS) archive for high resolution UV QSO spectra with sufficient signal-to-noise (S/N) to search and model the weak metal-line absorption features that may be signatures of the diffuse CGM gas of M31. Our search radius is within about 2 virial radii of M31. We adopt throughout a distance of 752 kpc and a virial radius of $R_{\rm vir} = 300$ kpc for M31 \citep{riess12,marel12}.\footnote{We emphasize that the exact definition of CGM and even $R_{\rm vir}$ is far from settled, as recently discussed by \citet{shull14}. For example, the overdensity $\Delta_{\rm vir}= 18 \pi^2 = 178 \approx 200$  of a collapsed object to estimate $R_{\rm vir}$ is often calculated assuming a ``top-hat'' model in an Einstein-de Sitter cosmology, but this value changes with the adopted cosmological models (e.g., \citealt{klypin02} adopted $\Delta_{\rm vir}  = 340$; see discussion in \citealt{shull14}). For ease of comparison with previous studies and in view of the distribution of our targets we choose for this paper $R_{\rm vir} = 300$ kpc for Andromeda. As we show in this paper, the gas that we associate with the CGM of M31 is at velocities consistent with the material being gravitationally bound to M31 (even beyond $R_{\rm vir}$), which implies that the  extended CGM gas of M31 is bound.} In  \S\ref{s-sample}, we describe in detail how the sample was selected, while in \S\ref{s-assoc} we present several arguments that point to an association to the diffuse CGM of M31 for the  absorption at $-300 \la v_{\rm LSR} \la -150$ \km\ observed along some of the sightlines in our sample . In \S\ref{s-prop}, we determine the properties of the M31 CGM, including its kinematics, ionization, covering fraction, and baryon and metal masses. We discuss the implications of our findings in \S\ref{s-disc} and finally summarize  our results in \S\ref{s-sum}.

\begin{figure*}[tbp]
\epsscale{1} 
\plotone{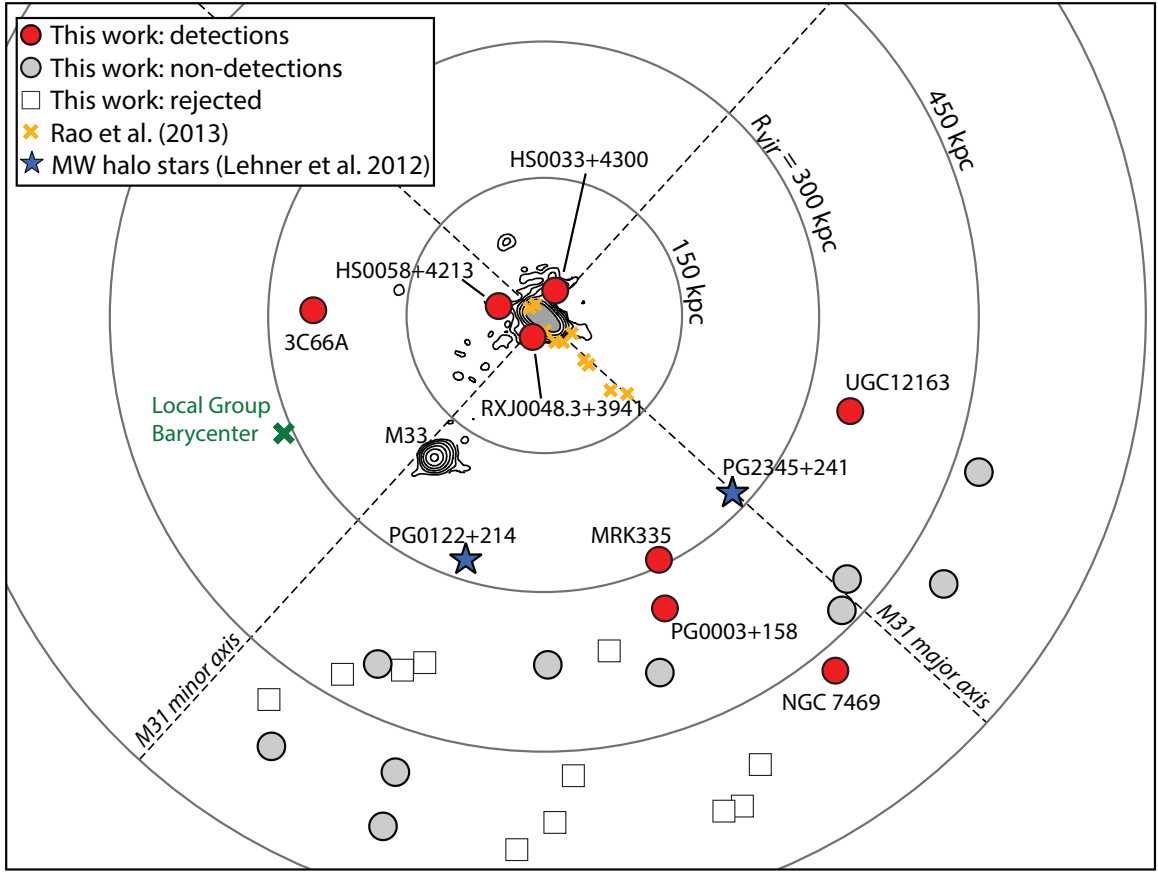}  
  \caption{Distribution of our targets in our adopted (circles) and rejected (squares) samples used to probe the CGM of M31 (right ascension increases from right to left, declination increases from bottom to top, see Table~\ref{t-data}). All these these sightlines were observed with COS G130M and/or G160M and some were also observed with \fuse. Targets with UV absorption at LSR velocities $-300 \le v_{\rm LSR} \le -150$ \km\ associated with the CGM of M31 are shown in red. Note that the sightline HS0058+4213 is near the $0.2 R_{\rm vir}$ boundary. Overplotted is the \hit\ 21-cm emission map around M31 adapted from \citet{braun04} where the lowest contour has $\log N_{\rm HI} = 17.5$ in the $40 \arcmin$ beam, with 0.5 dex increment between contours (this provides in our opinion a better representation of the \hit\ distribution around M31 according to recent GBT observations, see \citealt{lockman12,wolfe13}). The yellow crosses show for reference the targets from the COS G140L M31 program where low ions (e.g., \mgiit)  were detected only within the \hit\ disk contour (i.e., $R \le 32$ kpc, see \citealt{rao13}). The two MW stars in blue are distant halo stars, allowing us to determine that absorption at $v_{\rm LSR}\ga -170$ \km\ traces MW gas. We also indicate the position of the Local Group barycenter with the green cross. 
 \label{f-map}}
\end{figure*}

\section{Sample and Data Analysis}\label{s-sample}

In the \hst\ COS archive at the Mikulski Archive for Space Telescopes (MAST), we identify 31 targets at projected distances $R <2 R_{\rm vir}$ from M31 that were observed with the COS G130M and/or G160M gratings ($R_\lambda \approx 16,000$). In Table~\ref{t-data}, we summarize the 18 sightlines that are included in the adopted sample and the 13 sightlines rejected for the main analysis of the M31 CGM absorption.  The two chief reasons for excluding sightlines are either the signal-to-noise (S/N) level is too low (S/N\,$\la 4$) or there is a clear contamination from the MS or unrelated absorbers; the exact reasons for the exclusion of each sightline are given in the table footnotes and in \S\ref{s-assoc}, where we discuss the absorption velocity structure toward M31. In Fig.~\ref{f-map}, we show the distribution of the targets in our sample overplotted with the \hi\ contours from \citet{braun04}.    The three inner targets are also part of another paper that will consider in more detail the relation (if any) between the gas and stellar halos of M31 (C. Danforth et al. 2015, in prep.); only these targets were initially from a program aimed to study specifically the halo of M31 (\hst\ program 11632, PI: R. Rich). 

\begin{deluxetable*}{lccccccc}[t]
\tablewidth{0pc}
\tablecaption{Sample of COS G130M-G160M Targets Piercing the CGM of M31  \label{t-data}}
\tabletypesize{\footnotesize}
\tablehead{\colhead{Target} &
\colhead{RA} &
\colhead{DEC} &
\colhead{$z_{\rm em}$} &
\colhead{$R$} &
\colhead{COS} & 
\colhead{S/N} \\
\colhead{} & 
\colhead{($\degr$)} &
\colhead{($\degr$)} &
\colhead{} & 
\colhead{(kpc)} & 
\colhead{} & 
\colhead{} & 
\colhead{Note} 
}
\startdata
\cutinhead{Adopted Sample}     
 RXJ0048.3+3941	 	    &  12.08	 &   $+$39.69     &   0.134   &     25   &   G130M--G160M     &  23     &  1 \\ 
 HS0033+4300		    &  9.096	 &   $+$43.28     &   0.120   &     31   &   G130M--G160M     &   5     &  1 \\ 
 HS0058+4213		    &  15.38	 &   $+$42.49     &   0.190   &     49   &   G130M--G160M     &   6     &  1 \\ 
 3C66A  		    &  35.67	 &   $+$43.04     &   0.340   &    239   &   G130M--G160M     &  16     &  1 \\ 
 MRK335 		    &  1.58	 &   $+$20.20     &   0.034   &    287   &   G130M--G160M     &  20     &  1 \\ 
 PG0003+158		    &  1.50	 &   $+$16.16     &   0.450   &    334   &   G130M--G160M     &  16     &  1 \\
 UGC12163		    &  340.66	 &   $+$29.73     &   0.025   &    340   &   G130M--G160M     &   9     &  1 \\ 
 MRK1502		    &  13.40	 &   $+$12.69     &   0.061   &    361   &   G130M	      &   7     &  1 \\ 
 MRK1501		    &   2.63     & $+$10.984      &   0.090   &    389   &   G130M	      &   9     &  1\\ 
 SDSSJ015952.95+134554.3    &  29.97	 &   $+$13.77     &   0.504   &    401   &   G130M	      &   7     &  1 \\ 
 3C454.3		    &  343.49	 &   $+$16.15     &   0.859   &    424   &   G130M--G160M     &   5     &  1 \\
 SDSSJ225738.20+134045.0    &  344.41	 &   $+$13.68     &   0.594   &    440   &   G130M--G160M     &   6     &  1 \\ 
 NGC7469		    &  345.82	 &    $+$8.87     &   0.016   &    475   &   G130M--G160M     &  24     &  1 \\ 
 HS2154+2228		    & 329.20     & $+$22.713      &   1.290   &    476   &   G160M	      &   7     &  1  \\ 
 PHL1226		    &  28.62	 &    $+$4.81     &   0.404   &    482   &   G130M	      &   8     &  1 \\ 
 MRK304 		    &  334.30	 &   $+$14.24     &   0.067   &    499   &   G130M--G160M     &  20     &  1 \\ 
 MRK595 		    &  40.40	 &    $+$7.19     &   0.027   &    514   &   G130M 	      &   8     &  1 \\ 
 MRK1014		    &  29.96	 &    $+$0.40     &   0.164   &    527   &   G130M	      &  22     &  1 \\
 \cutinhead{Rejected Sample}     
  PG0026+129		      &   7.31     & $+$13.277      &   0.142    &    355   &   G130M	        &  12   &  2   \\ 
  SDSSJ014143.20+134032.0     &  25.43     & $+$13.685      &   0.045    &    381   &	G130M		&   3   &  3 \\ 
  IRAS01477+1254	      &  27.62     & $+$13.159      &   0.147    &    396   &	G130M--G160M	&   3   &  3   \\ 
  SDSSJ021348.53+125951.4     &  33.45     & $+$13.007      &   0.228    &    427   &	G130M--G160M	&   2   &  3   \\ 
  PG0044+030		      &  11.77     &  $+$3.331      &   0.624    &    462   &	G130M		&   6   &  4   \\ 
  4C10.08		      &  40.62     & $+$11.026      &   2.680    &    484   &	G130M		&   0   &  3   \\ 
  LBQS0052-0038		      &  13.86     &  $-$0.363      &   0.167    &    501   &	G130M--G160M	&   4   &  3   \\ 
  NGC7714		      & 354.06     &  $+$2.165      &   0.009    &    502   &	G130M--G160M	&   8   &  5   \\ 
  PG2349-014		      & 357.98     &  $-$1.153	    &   0.174    &    522   &	G130M		&   7   &  6  \\ 
  LBQS0107-0232 	      &  17.56     &  $-$2.282	    &   0.728    &    522   &	G160M		&   9   &  7  \\ 
  LBQS0107-0233 	      &  17.57     &  $-$2.314	    &   0.956    &    523   &	G130M--G160M	&   9   &  8  \\ 
  LBQS0107-0235 	      &  17.56     &  $-$2.331	    &   0.960    &    523   &	G130M--G160M	&   9   &  8  \\ 
  SDSSJ234500.43-005936.0     & 356.25     &  $-$0.993	    &   0.789    &    525   &	G130M--G160M	&   5   &  9  
\enddata
\tablecomments{
1: All the targets in the adopted sample satisfied the following criteria:  i) S/N\,$\ga 5$ in at least one transition, ii) there is no obvious contamination in the $-300 \la v_{\rm LSR}\la -150$ \km\ component from the MS in the direction of the target.    
2: Weak absorption features are observed between $-300$ and $-180$ \km\ in the profiles of \ciit, \siiit, \siiiit, and \siivt, but there is no match in the velocity centroids between all the observed transitions (including for the \siivt\ doublet), implying that there is some contamination from unrelated absorbers for several of the observed transitions. 
3: These data have too low S/N. 
4: Two absorption components at $-290$ and $-210$ \km, but at this RA, DEC, both are more likely to have a MS origin. 
5. Smeared absorption lines owing to the extended background object.
6: This sightline pierces the MS and has a MS component centered at $-300$ \km\ that extends from $-400$ to $-190$ \km.
7: Only \civt\ is available and both transitions of the doublet are contaminated. 
8: Absorption is observed at $-250$ \km\ in \siiiit\ and \civt, but at this RA, DEC, it is likely from the MS (this sightline is about 20\degr\ off the MS body). 
9: This sightline pierces the MS and has a MS component centered at $-250$ \km\ that extends from $-350$ to $-150$ \km.
}
\end{deluxetable*}

We also searched the {\it Far Ultraviolet Spectroscopic Explorer}\ (\fuse, $R_\lambda \approx 15,000$) archive to complement the COS data with the \ovi\ absorption. The following targets have \fuse\ observations with adequate S/N (i.e., $\ga 5$): RXJ0048.3+3941, MRK335, UGC12163, and NGC7469. We, however, do not consider \fuse\ data with no COS observation because the available UV transitions in the far-UV spectrum (\ovi, \cii, \ciii,  \siii) are either too weak or too contaminated to allow us to reliably identify individual velocity components in their absorption profiles. 

Information on the design and performance of COS and \fuse\ can be found in \citet{green12} and \citet{moos00}, respectively. Detailed information about the COS and \fuse\ data processing can be found in \citet{fox14} and \citet{wakker03}. In short, the exposures are aligned in wavelength space by matching the centroids of all the absorption lines visible in individual spectra. The coadded spectrum for each object is then shifted into the local standard of rest (LSR) velocity frame by matching the centroids of the Galactic \hi\ 21-cm emission in that direction from the LAB survey \citep{kalberla05} to the absorption of neutral and singly ionized species. We normalize the QSO continua with low order Legendre polynomials within about $\pm 1000$ \km\ from the absorption under consideration; typically polynomials with orders $m \le 4$ are fitted.  In Fig.~\ref{f-example}, we show an example of the normalized profiles of selected ions for RXJ0048.1+3941. In the figures of the Appendix, we show all the normalized profiles of  our selected targets against the LSR velocity.  The \ovi\ can be contaminated by H$_2$ absorption, and we remove this contamination following the method described in \citet{wakker06}. We show the H$_2$ model in the figures of the Appendix. Based on previous experience, this contamination can be removed fairly accurately with an uncertainty of about $\pm 0.1$ dex on the column density \citep{wakker03}. 

To search for M31 CGM absorption and to determine the properties of the CGM gas, we consider the following atomic and ionic and transitions: \oi\ $\lambda$1302,  \cii\ $\lambda$1334, \ciii\ $\lambda$977, \civ\ $\lambda$$\lambda$1548, 1550, \siii\ $\lambda$1193 (\siii\ $\lambda$1260 is blended with \sii\ $\lambda$1259 near the systemic velocity of M31), \siiii\ $\lambda$1206,  \siiv\ $\lambda$$\lambda$1393, 1402, as well as \ovi\ $\lambda$$\lambda$1031, 1037 for the four sightlines observed with \fuse. Other transitions are typically too weak to be detected and do not produce interesting limits.

We use the apparent optical depth (AOD) method described by \citet{savage91} to estimate the total column density and average velocity of the metal ions. The absorption profiles are converted into apparent column densities per unit velocity, $N_a(v) = 3.768\times 10^{14} \ln[F_c(v)/F_{\rm obs}(v)]/(f\lambda)$ cm$^{-2}$\,(\km)$^{-1}$, where $F_c(v)$ and $F_{\rm obs}(v)$ are the modeled continuum and observed fluxes as a function of velocity, respectively, $f$ is the oscillator strength of the transition and $\lambda$ is the wavelength in \AA. The atomic parameters are for the UV transitions from \citet{morton03}.  When no detection was observed, we estimated a 3$\sigma$ upper limit following the method described by \citet{lehner08}. The total column density was obtained by integrating over the absorption profile  $N = \int_{v_1}^{v_2} N_a(v)dv$ and the average velocity was determined from the first moment $\langle v \rangle = \int_{v_1}^{v_2} v N_a(v)dv/\int_{v_1}^{v_2} N_a(v)dv$.

We consider absorption over the velocities that we associate with the CGM of M31 (see the component in red in Fig.~\ref{f-example}); see \S\ref{s-assoc} for several arguments that support this association. In Table~\ref{t-ions}, we summarize the integration range for each sightline and the velocity and column density results for each species. We find a good agreement between the column densities estimated for the doublet transitions (\civ, \siiv), showing no indication of saturation or contamination. We note, however, that \cii\ and \siiii\ for the 3 targets at $R \le 50$ kpc ($R\le 0.2 R_{\rm vir}$) could be somewhat saturated owing the peak apparent optical depth being $\tau_a \ga 1$. In this table, we also list the average velocity (defined as $\langle v_{\rm all} \rangle$) based on all the reliable transitions for a given sightline. For the sightlines where no M31 absorption is found, we derive a $3\sigma$ upper using generally the velocity interval $[-280,-150]$  \km\ to mimic the velocity range of the absorption attributed to the M31 CGM observed within $0.2 R_{\rm vir}$. 

\begin{deluxetable}{lccc}
\tablewidth{0pc}
\tablecaption{Kinematics and Column Densities  \label{t-ions}}
\tabletypesize{\footnotesize}
\tablehead{\colhead{Species} & \colhead{$(v_1,v_2)$} & \colhead{$\langle v \rangle$} &\colhead{$\log N$} \\
\colhead{} & \colhead{(\km)} & \colhead{(\km)} &\colhead{[cm$^{-2}$]}
}
\cutinhead{ RXJ0048.3+3941, $R = 25$ kpc, $\langle v_{\rm all} \rangle = -213.1 \pm 8.7 $ \km}     
        \oit\ $\lambda$1302 & $  -306, -150 $ & 	     \nodata   & $ <	   13.63   $ \\
       \ciit\ $\lambda$1334 & $  -306, -150 $ & $  -188.7  :	    $  & $14.57  :	   $ \\
      \siiit\ $\lambda$1193 & $  -306, -150 $ & $  -197.4  \pm  2.0 $  & $13.43  \pm 0.02  $ \\
      \ciiit\ $\lambda$977  & $  -306, -150 $ & $  -220.7 :         $  & $>14.08 	   $ \\
     \siiiit\ $\lambda$1206 & $  -306, -150 $ & $  -215.5  \pm  0.8 $  & $13.42  \pm 0.01  $ \\
      \siivt\ $\lambda$1393 & $  -306, -150 $ & $  -212.9  \pm  1.3 $  & $13.38  \pm 0.01  $ \\
      \siivt\ $\lambda$1402 & $  -306, -150 $ & $  -211.3  \pm  2.6 $  & $13.41  \pm 0.03  $ \\
       \civt\ $\lambda$1548 & $  -306, -150 $ & $  -218.8  \pm  1.1 $  & $14.08  \pm 0.02  $ \\
       \civt\ $\lambda$1550 & $  -306, -150 $ & $  -222.7  \pm  2.0 $  & $14.11  \pm 0.02  $ \\
       \ovit\ $\lambda$1031 & $  -306, -130 $ & $  -209.4  \pm  5.0 $  & $14.38  \pm 0.12  $ \\
\cutinhead{HS0033+4300, $R = 31$ kpc, $\langle v_{\rm all} \rangle = -204.1 \pm 2.0 $ \km}
        \oit\ $\lambda$1302 & $  -280, -150 $ & 	     \nodata   & $ <	   14.26   $ \\
       \ciit\ $\lambda$1334 & $  -280, -150 $ & $  -204.1  \pm  5.9 $  & $14.34  :	   $ \\
      \siiit\ $\lambda$1193 & $  -280, -150 $ & $  -223.8  :	    $  & $13.36  \pm 0.18  $ \\
     \siiiit\ $\lambda$1206 & $  -280, -150 $ & $  -201.2  \pm  6.4 $  & $13.31  \pm 0.10  $ \\
      \siivt\ $\lambda$1402 & $  -280, -150 $ & $  -206.8  \pm  6.7 $  & $13.26  \pm 0.09  $ \\
       \civt\ $\lambda$1548 & $  -280, -150 $ & $  -203.6  \pm  3.0 $  & $14.04  \pm 0.05  $ \\
       \civt\ $\lambda$1550 & $  -280, -150 $ & $  -204.6  \pm  5.5 $  & $14.07  \pm 0.08  $ \\
\cutinhead{  HS0058+4213, $R = 48$ kpc, $\langle v_{\rm all} \rangle = -211.1 \pm 3.9 $ \km}
        \oit\ $\lambda$1302 & $  -280, -150 $ & 	     \nodata   & $ <	   14.02   $ \\
       \ciit\ $\lambda$1334 & $  -280, -150 $ & $  -214.1  \pm  2.9 $  & $14.41  \pm 0.06  $ \\
      \siiit\ $\lambda$1193 & $  -280, -150 $ & $  -217.3  \pm  5.9 $  & $13.45  \pm 0.08  $ \\
     \siiiit\ $\lambda$1206 & $  -280, -150 $ & $  -210.5  \pm  3.2 $  & $13.44  \pm 0.08  $ \\
      \siivt\ $\lambda$1393 & $  -280, -150 $ & $  -207.4  \pm  5.1 $  & $13.42  \pm 0.07  $ \\
      \siivt\ $\lambda$1402 & $  -280, -150 $ & $  -212.4  \pm  8.7 $  & $13.43  \pm 0.12  $ \\
       \civt\ $\lambda$1548 & $  -280, -150 $ & $  -210.5  \pm  4.3 $  & $13.95  \pm 0.06  $ \\
       \civt\ $\lambda$1550 & $  -280, -150 $ & $  -205.8  \pm  7.6 $  & $14.00  \pm 0.10  $ \\
\cutinhead{3C66A, $R = 239$ kpc, $\langle v_{\rm all} \rangle =  -256.0 \pm  4.9 $ \km}
        \oit\ $\lambda$1302& $  -321, -214 $ &  	    \nodata    & $ <	 13.54     $ \\
       \ciit\ $\lambda$1334& $  -321, -214 $ &  	    \nodata    & $ <	 13.10     $ \\
      \siiit\ $\lambda$1193& $  -321, -214 $ &  	    \nodata    & $ <	 12.54     $ \\
     \siiiit\ $\lambda$1206& $  -321, -214 $ & $  -256.0  \pm  4.9 $   & $\le 12.48  \pm 0.07  $ \\
      \siivt\ $\lambda$1393& $  -321, -214 $ &  	    \nodata    & $ <	 13.04     $ \\
       \civt\ $\lambda$1548& $  -321, -214 $ &  	    \nodata    & $ <	 13.02     $ \\
\cutinhead{MRK335, $R = 287$ kpc, $\langle v_{\rm all} \rangle = -240.1 \pm 5.2 $ \km}
        \oit\ $\lambda$1302 & $  -275, -180 $ & 	     \nodata   & $ <	  13.43   $ \\
       \ciit\ $\lambda$1334 & $  -275, -180 $ & 	     \nodata   & $ <	  12.98   $ \\
      \ciiit\ $\lambda$977  & $  -275, -180 $ & $  -222.4  :	    $  & $>14.09	  $ \\
     \siiit\  $\lambda$1193 & $  -275, -180 $ & 	     \nodata   & $ <	  12.42   $ \\
     \siiiit\ $\lambda$1206 & $  -275, -150 $ & $  -238.5  \pm  5.6 $  & $12.41  \pm 0.06 $ \\
      \siivt\ $\lambda$1393 & $  -275, -180 $ & $  -241.1  \pm  6.3 $  & $12.62  \pm 0.09 $ \\
      \siivt\ $\lambda$1402 & $  -275, -150 $ & $  -247.9  \pm 15.5 $  & $12.69  \pm 0.15 $ \\
       \civt\ $\lambda$1548 & $  -275, -180 $ & $  -233.7  \pm  2.5 $  & $13.48  \pm 0.04 $ \\
       \civt\ $\lambda$1550 & $  -275, -180 $ & $  -239.1  \pm  5.6 $  & $13.40  \pm 0.10 $ \\
       \ovit\ $\lambda$1031 & $  -275, -180 $ & $  -222.0  \pm  1.0 $  & $14.10  \pm 0.13 $ \\
\cutinhead{PG0003+158, $R = 334$ kpc, $\langle v_{\rm all} \rangle = -232.8 \pm 4.4 $ \km}
        \oit\ $\lambda$1302 & $  -275, -186 $ & 	     \nodata   & $ <	   13.59   $ \\
     \siiit\  $\lambda$1193 & $  -275, -186 $ & 	     \nodata   & $ <	   12.48   $ \\
     \siiiit\ $\lambda$1206 & $  -275, -186 $ & $  -235.6  \pm  2.4 $  & $12.63  \pm 0.04  $ \\
      \siivt\ $\lambda$1393 & $  -275, -186 $ & $  -226.5  \pm  7.2 $  & $12.51  \pm 0.13  $ \\
       \civt\ $\lambda$1548 & $  -275, -186 $ & $  -235.8  \pm  2.9 $  & $13.28  \pm 0.06  $ \\
       \civt\ $\lambda$1550 & $  -275, -186 $ & $  -233.4  \pm  6.6 $  & $13.22  \pm 0.16  $ \\
\cutinhead{UGC12163, $R = 339$ kpc, $\langle v_{\rm all} \rangle = -274.3 \pm 5.1 $ \km}
        \oit\ $\lambda$1302 & $  -280, -150 $ & 	     \nodata   & $ <	  14.08   $ \\
       \ciit\ $\lambda$1334 & $  -280, -150 $ & 	     \nodata   & $ <	  13.53   $ \\
     \siiit\  $\lambda$1193 & $  -280, -150 $ & 	     \nodata   & $ <	  13.00   $ \\
     \siiiit\ $\lambda$1206 & $  -280, -150 $ & 	     \nodata   & $ <	  12.54   $ \\
      \siivt\ $\lambda$1393 & $  -280, -150 $ & 	     \nodata   & $ <	  12.95   $ \\
       \civt\ $\lambda$1548 & $  -280, -150 $ & 	     \nodata   & $ <	  13.35   $ \\
       \ovit\ $\lambda$1031 & $  -320, -200 $ & $  -274.3  \pm  5.1 $  & $\le 14.20  \pm 0.17 $ \\
\cutinhead{MRK1502, $R = 361$ kpc}
       \ciit\ $\lambda$1334 & $  -280, -150 $ & 	     \nodata   & $ <	   13.48   $ \\
     \siiit\  $\lambda$1193 & $  -280, -150 $ & 	     \nodata   & $ <	   12.85   $ \\
     \siiiit\ $\lambda$1206 & $  -280, -150 $ & 	     \nodata   & $ <	   12.57   $ \\
      \siivt\ $\lambda$1393 & $  -280, -150 $ & 	     \nodata   & $ <	   12.97   $ \\
\cutinhead{MRK1501, $R = 389$ kpc}
       \ciit\ $\lambda$1334 & $  -280, -150 $ & 	     \nodata   & $ <	   13.33   $ \\
\cutinhead{SDSSJ015952.95+134554.3, $R = 401$ kpc}
       \ciit\ $\lambda$1334 & $  -280, -150 $ & 	     \nodata   & $ <	   13.56   $ \\
     \siiit\  $\lambda$1193 & $  -280, -150 $ & 	     \nodata   & $ <	   12.96   $ \\
     \siiiit\ $\lambda$1206 & $  -270, -220 $ & 	     \nodata   & $ <	   12.29   $ \\
      \siivt\ $\lambda$1393 & $  -280, -150 $ & 	     \nodata   & $ <	   12.93   $ \\
\cutinhead{3C454.3, $R = 424$ kpc}
       \ciit\ $\lambda$1334 & $  -280, -150 $ & 	     \nodata   & $ <	   13.70   $ \\
     \siiit\  $\lambda$1193 & $  -280, -150 $ & 	     \nodata   & $ <	   13.19   $ \\
     \siiiit\ $\lambda$1206 & $  -280, -150 $ & 	     \nodata   & $ <	   12.79   $ \\
      \siivt\ $\lambda$1393 & $  -280, -150 $ & 	     \nodata   & $ <	   13.07   $ \\
       \civt\ $\lambda$1548 & $  -280, -150 $ & 	     \nodata   & $ <	   13.29   $ \\
\cutinhead{SDSSJ225738.20+134045.0, $R = 440$ kpc}
       \ciit\ $\lambda$1334 & $  -280, -150 $ & 	     \nodata   & $ <	   13.64   $ \\
     \siiit\  $\lambda$1193 & $  -280, -150 $ & 	     \nodata   & $ <	   13.14   $ \\
     \siiiit\ $\lambda$1206 & $  -280, -150 $ & 	     \nodata   & $ <	   12.66   $ \\
      \siivt\ $\lambda$1393 & $  -280, -150 $ & 	     \nodata   & $ <	   13.06   $ \\
       \civt\ $\lambda$1550 & $  -280, -150 $ & 	     \nodata   & $ <	   13.71   $ \\
\cutinhead{NGC7469, $R = 475$ kpc, $\langle v_{\rm all} \rangle = -239.4 \pm 3.3 $ \km}
    	\oit\ $\lambda$1302 & $  -260, -205 $ & 	     \nodata   & $ <	   13.28   $ \\
       \ciit\ $\lambda$1334 & $  -260, -205 $ & $  -239.3  \pm  2.6 $  & $13.16  \pm 0.07  $ \\
      \siiit\ $\lambda$1193 & $  -260, -205 $ & 	     \nodata   & $ <	   12.23   $ \\
     \siiiit\ $\lambda$1206 & $  -260, -205 $ & $  -241.9  \pm  1.0 $  & $12.60  \pm 0.03  $ \\
      \siivt\ $\lambda$1393 & $  -260, -205 $ & $  -234.6  \pm  1.7 $  & $12.74  \pm 0.05  $ \\
      \siivt\ $\lambda$1402 & $  -260, -205 $ & $  -243.6  \pm  4.0 $  & $12.69  \pm 0.10  $ \\
       \civt\ $\lambda$1548 & $  -260, -205 $ & $  -237.0  \pm  1.9 $  & $13.20  \pm 0.05  $ \\
       \civt\ $\lambda$1550 & $  -260, -205 $ & $  -240.0  \pm  2.8 $  & $13.27  \pm 0.08  $ \\
       \ovit\ $\lambda$1031 & $  -260, -205 $ & $  -230.3  \pm  2.4 $  & $13.87  \pm 0.15  $ \\
\cutinhead{HS2154+2228, $R = 476$ kpc}
       \civt\ $\lambda$1550 & $  -270, -150 $ & 	     \nodata   & $ <	   13.31   $ \\
\cutinhead{PHL1226, $R = 482$ kpc}
       \ciit\ $\lambda$1334 & $  -280, -150 $ & 	     \nodata   & $ <	   13.51   $ \\
     \siiit\  $\lambda$1193 & $  -280, -150 $ & 	     \nodata   & $ <	   12.92   $ \\
     \siiiit\ $\lambda$1206 & $  -280, -150 $ & 	     \nodata   & $ <	   12.46   $ \\
      \siivt\ $\lambda$1402 & $  -280, -150 $ & 	     \nodata   & $ <	   13.14   $ \\
\cutinhead{MRK304, $R = 495$ kpc}
       \ciit\ $\lambda$1334 & $  -280, -150 $ & 	     \nodata   & $ <	   13.02   $ \\
     \siiit\  $\lambda$1193 & $  -280, -150 $ & 	     \nodata   & $ <	   12.57   $ \\
     \siiiit\ $\lambda$1206 & $  -280, -150 $ & 	     \nodata   & $ <	   12.14   $ \\
      \siivt\ $\lambda$1393 & $  -280, -150 $ & 	     \nodata   & $ <	   12.47   $ \\
       \civt\ $\lambda$1548 & $  -280, -150 $ & 	     \nodata   & $ <	   13.03   $ \\
\cutinhead{MRK595, $R = 514$ kpc}
       \ciit\ $\lambda$1334 & $  -310, -180 $ & 	     \nodata   & $ <	   13.47   $ \\
     \siiit\  $\lambda$1193 & $  -310, -180 $ & 	     \nodata   & $ <	   12.95   $ \\
     \siiiit\ $\lambda$1206 & $  -310, -180 $ & 	     \nodata   & $ <	   12.51   $ \\
      \siivt\ $\lambda$1393 & $  -310, -180 $ & 	     \nodata   & $ <	   12.86   $ \\
\cutinhead{MRK1014, $R = 527$ kpc}
       \ciit\ $\lambda$1334 & $  -280, -150 $ & 	     \nodata   & $ <	   13.33   $ \\
     \siiit\  $\lambda$1193 & $  -280, -150 $ & 	     \nodata   & $ <	   12.63   $ \\
     \siiiit\ $\lambda$1206 & $  -280, -150 $ & 	     \nodata   & $ <	   12.18   $ \\
      \siivt\ $\lambda$1402 & $  -280, -150 $ & 	     \nodata   & $ <	   12.84   $ 
\enddata
\tablecomments{All the velocities are in the LSR frame. The velocities $v_1$ and $v_2$ are the velocities used for the integration of the AOD profiles. A colon means that the estimate value is uncertain owing to blending. A ``$<$ '' indicates that no absorption is detected and a 3$\sigma$ upper limit is reported.  A ``$>$ '' sign is lower limit because the absorption reaches the zero-flux level. A ``$\le $'' sign indicates that only  one transition of one ion  is detected; this sign emphasizes that the line could be contaminated. For the \ovit, the errors are dominated by systematics from the correction of the H$_2$ contamination. When more than one transition is detected, $\langle v_{\rm all}\rangle$ is the average velocity between all the ions (observed with COS) where the detection is not blended or uncertain.}
\end{deluxetable}

\section{Absorption Velocity Structure and Association with M31}\label{s-assoc}
\begin{figure}[tbp]
\epsscale{1.1} 
\plotone{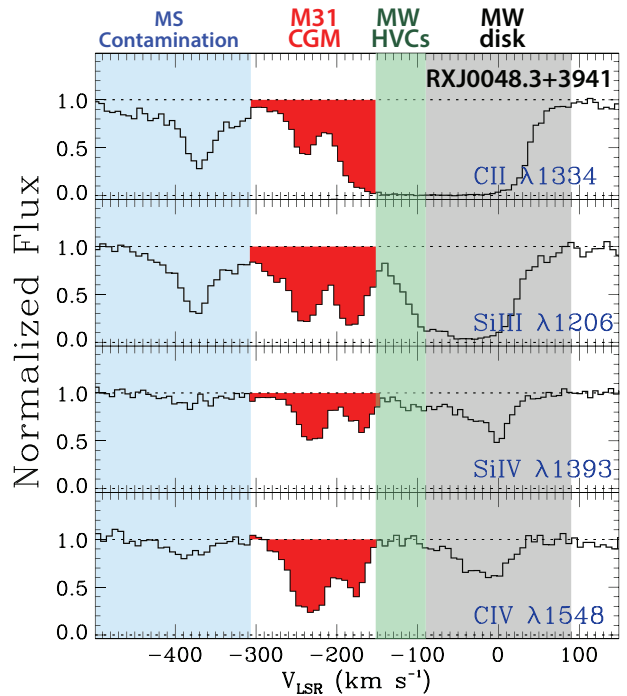}  
  \caption{Normalized profiles of  RXJ0048.1+3941 ($R = 25$ kpc) as a function of the LSR velocity, where we highlight the origin of each component, the MS, the CGM of M31, and the MW disk and HVCs based on the arguments presented in \S\ref{s-assoc}. All the species for each selected sightline studied in this work are shown in the Appendix.
 \label{f-example}}
\end{figure}

In the general direction of M31, the absorption seen in the QSO spectra is complex, with absorption possibly arising from the MW thick disk and halo, MS, M31 CGM, Local Group (LG), or unrelated absorbers at higher redshifts, $z$. Unrelated absorbers are ruled out because for all but two sightlines we use several species and different transitions of the same species to identify higher redshift absorbers. In Fig.~\ref{f-example}, we show an example of the normalized profiles of \cii, \siii, \siiii\, and \civ\ for RXJ0048.1+3941 at a projected distance $R=25$ kpc from  M 31 (see the normalized spectra in the Appendix for the entire sample). This sightline shows absorption in several complexes stretching from $-490$ \km\ to $+90$ \km.  As we show below the absorption at $v_{\rm LSR}\ga -150$ \km\ is associated with the MW disk and halo; when detected, the absorption at $-430 \la v_{\rm LSR}\la -300$ \km\ is mostly associated with the MS, while we associate the $-300 \la v_{\rm LSR}\la -150$ \km\ component with the CGM of M31.

\subsection{MW: absorption at  $v_{\rm LSR}\ga -150$ \km}\label{s-mw}

Both the MW thick disk and halo absorption can be readily identified in the direction of M31. Toward all the sightlines, saturated absorption is observed in \cii, \siii, and \siiii\ between about $-90$ and $+50$ \km. This absorption arises principally from the MW thick disk based on the Galactic disk rotation curve \citep{clemens85} and countless studies of the MW in UV absorption \citep[e.g.,][]{shull09,lehner12,wakker12}. 

 Toward many sightlines in our sample, absorption is also observed at velocities between about $-170$ and $-90$ \km. This absorption is from the MW iHVCs that are at distances of $d\la 10$--15 kpc from the sun. We come to this conclusion for two reasons.  First \citet{lehner11} and \citet{lehner12} statistically constrained the distance of the iHVC population, showing that absorption at $90 \le |v_{\rm LSR}|\la 170$ \km\ is found typically in the lower halo of the MW, as it is observed at about the same rate toward MW halo stars and extragalactic AGN. Thus the gas seen in absorption toward M31 at $-170 \la v_{\rm LSR}\la -90$ \km\ corresponds to MW gas within 5--15 kpc of the sun. Second we show in Fig.~\ref{f-map} the locations of two MW halo stars near on the sky to M31 and M33 that were observed with COS (\hst\ programs 12982 and 11592; PI: Lehner). These two stars are PG0122+214 ($d =9.6$ kpc), where HVC absorption is observed at  $v_{\rm LSR} \simeq -160$ \km\ \citep[see][]{lehner12}, and PG2345+241 ($d=4.9$ kpc, near the MRK335 sightline; see Fig.~\ref{f-map}), where HVC absorption is seen at $v_{\rm LSR} \simeq -120$ \km. In both cases, the HVC absorption is very weak and detected only in \siii\ and \cii, not in \civ\ or \siiv, which are typically observed at $-300 \la v_{\rm LSR}\la -150$ \km\ (\siiii\ cannot be used in stellar spectra owing to the strong damping wings of \lya). Thus gas at $-150 \la v_{\rm LSR} \la -90$ \km\ is relatively close to the sun in this direction, and we therefore assign any absorption in our sample with velocity components centered on $v_{\rm LSR}\ga -150$ \km\ to the MW disk and its halo. 

\subsection{Identifying M31 CGM Gas}
Several of our sightlines show absorption at $v_{\rm LSR}\la -170$ \km. Gas at these velocities has not been seen toward halo stars at $d\la20$ kpc from the sun (see \S\ref{s-mw}). We identify the absorption at $-300 \la v_{\rm LSR}\la -150$ \km\ with the CGM of M31 for the following reasons:
\begin{enumerate}
\item These velocities differ by more than 200 \km\ for some of the targets from the expected velocities of the MS in this direction (\S \ref{s-velm31}). 
\item These velocities are similar to the velocities of the M31 dwarf satellites (\S \ref{s-velm31}). 
\item There is a non-trivial relationship between the column densities of this component and the projected distances from M31 that only exists for gas in this velocity range (\S \ref{s-nvsr}). 
\item The detection rate of this component is high within a projected distance of 300 kpc (the virial radius) from M31 and plummets beyond 300 kpc, another property only observed for this velocity component (\S \ref{s-nvsr}). 
\item The \hi\ emission observations have detected clouds in the M31 CGM at similar LSR velocities at projected distances $\la 100$ kpc (\S \ref{s-hicgm}).
\end{enumerate}
Each of these independent reasons alone points toward a M31 CGM origin. Combining all these observed properties of the absorption at $-300 \la v_{\rm LSR}\la -150$ \km\ allows us to associate with very little doubt this population of clouds with the CGM of M31. 

\begin{figure}[tbp]
\epsscale{1.2} 
\plotone{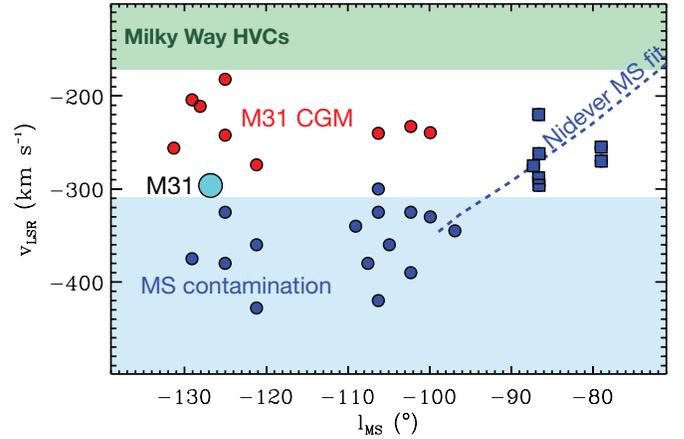}  
  \caption{LSR velocity of the individual components centered at  $v_{\rm LSR}<-170$ \km\ identified in the spectra of the targets summarized in Table~\ref{t-data}. The red and blue circles are from the adopted sample, while the blue squares are from the rejected sample.  The cyan circle represents the systemic velocity of M31 at the M31 location. The dashed line represents the average velocities LSR velocities of \hit\ estimated from the results by \citet{nidever08}. \label{f-vlms}}
\end{figure}

\subsubsection{Velocities}\label{s-velm31}
\noindent
The MS extends over this part of the sky.  In order to understand its potential signature in our observations, it is useful to plot the LSR velocity as a function of the MS longitude, $l_{\rm MS}$ (see \citealt{nidever08} for the definition of the Magellanic Stream coordinate) since  \citet{nidever08} found a tight relationship between the LSR velocity of the MS seen in \hi\ emission and $l_{\rm MS}$.  In Fig.~\ref{f-vlms},  the dashed line represents the average \hi\ LSR velocity of the MS with $l_{\rm MS}$  determined from Fig.~9 of \citet{nidever08}, which has a very small scatter in the velocity ($\sim 20$--30 \km) at these $l_{\rm MS}$; there is no detection in \hi\ emission (with the LAB sensitivity) beyond $ l_{\rm MS}\la -100\degr$. In each spectrum, we estimate the average velocities of the individual components in the velocity ranges  $-450 \la v_{\rm LSR} \la -300$ \km\ and $-300  \la v_{\rm LSR} \la -150$ \km, which are represented by the blue and red circles in Fig.~\ref{f-vlms}.  We also show in this figure with blue squares the data from the rejected sample sightlines for individual components at $v_{\rm LSR}\la -170$ \km.

The blue squares are all at $ l_{\rm MS}\ge -90\degr$ where the MS is still observed in \hi\ emission and where the expected MS are in the velocity range at $-300  \la v_{\rm LSR} \la -200$ \km. That apparent association with the MS at $-300  \la v_{\rm LSR} \la -200$ \km\ led us to reject them from the sample, so it is not surprising that the blue squares lie quite close to the dashed line.  At $-110\degr \le l_{\rm MS}\le -95\degr$, the blue circles are scattered around the general continuation of the dashed line. These  mostly likely trace gas from the MS that extends beyond the dense \hi\  body of the MS. 

The  $-130\degr \le l_{\rm MS}\le -110 \degr$ longitudes are more than $30\degr$--$40\degr$ from the MS seen in \hi\ 21-cm emission. However, \citet{sembach03} and \citet{fox14} found UV absorption at the expected  MS velocities extending $30\degr$ from the body of the MS tip seen in \hi\ emission and concluded that this gas represents a diffuse extension of the MS seen only in metal line absorption.  \citet{braun04} also noted extended \hi\ emission filaments at the level $\log N_{\rm HI}>17.3$ (i.e., a factor of 10 or more deeper than previous \hi\ surveys) for velocities $-430\la v_{\rm LSR}\la -330$ \km\ that align with the expected extension of the MS (see their Figs.~3 and 5).\footnote{The apparent smooth structures seen in Fig.~3 of \citet{braun04} are, however, questionable based on follow-up higher angular resolution GBT observations that reveal at least for the gas between M31 and M33 that the \hit\ emission is typically resolved into cloud-like structure the size of a GBT beam-width or smaller rather than elongated filamentary structure, and at much higher column densities \citep{lockman12,wolfe13}.} Based on these studies, it is  likely that the  blue circles in Fig.~\ref{f-vlms} at $-130\degr \le l_{\rm MS}\le -120 \degr$ trace predominantly the MS, extending  the tip of the MS observed in \hi\ emission by more than $30\degr$. If they are part of the MS, the velocity of the MS appears to  flatten between $ l_{\rm MS}\sim -130 \degr$ and $-100 \degr$. All the blue components at $l_{\rm MS}\ga -110 \degr$ in Fig.~\ref{f-vlms} are within $\la 50$ \km\ of the expected MS velocities.  

On the other hand, the red circles at $-130\degr \le l_{\rm MS}\le -120 \degr$ are at velocities incompatible with a MS origin by more than 100--200 \km. At  $-110\degr \le l_{\rm MS}\le -95\degr$, they are about 100 \km\ from the expected MS velocities determined by the dashed line, but only 50 \km\ from some of the components marked as blue cirlces. Given their proximity to MS velocities, the origin of the component $ v_{\rm LSR} \sim -220$ \km\ toward MRK335, PG0003+158, and NGC7469 could potentially arise in the Stream. The first 2 sightlines have a MS latitude $\sim 12\degr$, while NGC7469 has $b_{\rm MS} \simeq -5\degr$, i.e., NGC7469 goes through some of the \hi\ emission from the MS (see Fig.~8 in \citealt{nidever08}), making NGC7469 the most uncertain sightline. NGC7469 is also the farther sightline from M31 with $R \simeq 1.6 R_{\rm vir}$. The other two sightlines lie at projected distances from M31 $R\approx R_{\rm vir}$. 

\begin{figure}[tbp]
\epsscale{1.2} 
\plotone{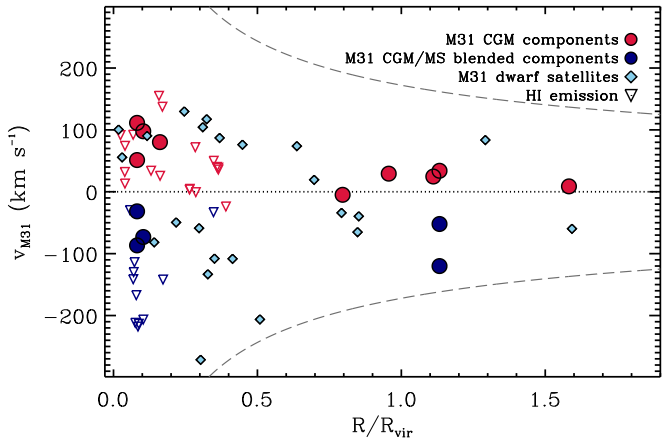}  
  \caption{Comparison of the M31-centric velocities of the gas components seen in UV absorption and \hit\ emission associated with the M31 CGM and the M31 galaxy satellites as a function of the projected distances from M31. For the blue circles, we only consider the data at $l_{\rm MS}\le -120 \degr$. The color coding for the circles is the same as those in Fig.~\ref{f-vlms}. The color coding of the 
upside down triangles follows that of the circles. The dashed curves show the escape velocity for a $10^{12}$ M$_\sun$ point mass.  The \hit\ emission data are at $R< 0.2 R_{\rm vir}$ from \citet{westmeier08} and at $R\ge 0.2 R_{\rm vir}$ from \citet{lockman12} and \citet{wolfe13}. The data for the galaxies associated with M31 were estimated from the parameters derived by \citet{mcconnachie12}.   \label{f-vm31}}
\end{figure}

From Fig.~\ref{f-vlms}, it is also noticeable how the red and blue circles together appear quasi-symmetrically distributed around the systemic velocity of M31 at $l_{\rm MS}\la -120 \degr$. The component at negative velocities relative to M31 is, however, not observed in all the inner region sightlines, only toward RXJ0048.1+3941 ($R=25$ kpc) and HS0033+4300  ($R = 31$ kpc) at $v_{\rm LSR} \simeq -380$ \km\ and farther away toward UGC12163 ($R=339$ kpc) at $v_{\rm LSR} \simeq -430$ \km\ (strong absorption, see figure in Appendix), and $-360$ \km\ (weak absorption). The MS most certainly dominates the absorption of these components because  1) the properties of the absorption are reminiscent of those observed toward sightlines known to probe the MS with  strong absorption in \cii\ and \siii\ as well as absorption in \siiii, \siiv, and \civ\ \citep[see, e.g.,][]{fox14}, and 2) there is no difference in the strength of the absorption at projected distances from M31 of 25--31 kpc and 339 kpc (see \S\ref{s-nvsr}). However, it is plausible that there is some absorption at these velocities from the M31 CGM gas that is strongly contaminated by the Magellanic Stream.

In order to learn more about the observed velocity distribution and determine if there is any similarity, we consider the dwarf galaxies that are located within the CGM of M31. \citet{mcconnachie12} recently summarized the properties of the Local group galaxies, finding 23 of them associated with M31. For this comparison, it is useful to consider a ``M31-centric" velocity defined as $v_{\rm M31} = v_{\rm GSR}({\rm Obj}) - v_{\rm GSR}({\rm M31})\,\cos(\theta)$, where $v_{\rm GSR} = v_{\rm LSR} + 220 \sin(l) \cos(b)$ is the velocity of in the Galactic standard-of-rest (GSR) frame and $\theta$ the angular separation of the galaxy or QSOs from M31. These velocities are very similar to the $v_{\rm M31}$ values listed in Table~2 of \citet{mcconnachie12}. Using the positions and radial velocities listed in Table~2 of his paper, we calculate the projected distances from M31 following the same method used for our sample and $v_{\rm M31} $ for each gas cloud in our sample and for each dwarf galaxy in the \citet{mcconnachie12} sample. We use the projected distances since we do not know the distance of the absorbers in our sample from the M31. However, we stress the membership of each dwarf galaxies with M31 is based on distances from the sun that are between 525 kpc and 920 kpc and within $<2R_{\rm vir}$ from M31 \citep[see for more detail][]{mcconnachie12}.

In Fig.~\ref{f-vm31}, the distribution of $v_{\rm M31}$ as a function of $R/R_{\rm vir}$ is shown for the dwarf satellites and the gas seen in absorption at $-300 \la v_{\rm LSR}\la -150$ \km\ (red circles) and $-420 \la v_{\rm LSR}\la -300$ \km\ with $-130\degr \le l_{\rm MS}\le -120 \degr$ (blue circles). The similar velocity distribution between the gas and dwarf satellites in the CGM of M31 is striking, adding more weight that the absorption features  at $-300 \la v_{\rm LSR}\la -150$ \km\ are indeed probing the  CGM gas of M31. This comparison also suggests that the absorption features at $-420 \la v_{\rm LSR}\la -300$ \km\ could arise in part from the CGM of M31. However, the other properties of this component are more similar to those seen in the MS; we therefore treat this possible component of the M31 CGM largely contaminated by the MS.

\subsubsection{Column Densities and Detection Rates  versus $R$}\label{s-nvsr}
\begin{figure}[tbp]
\epsscale{1.} 
\plotone{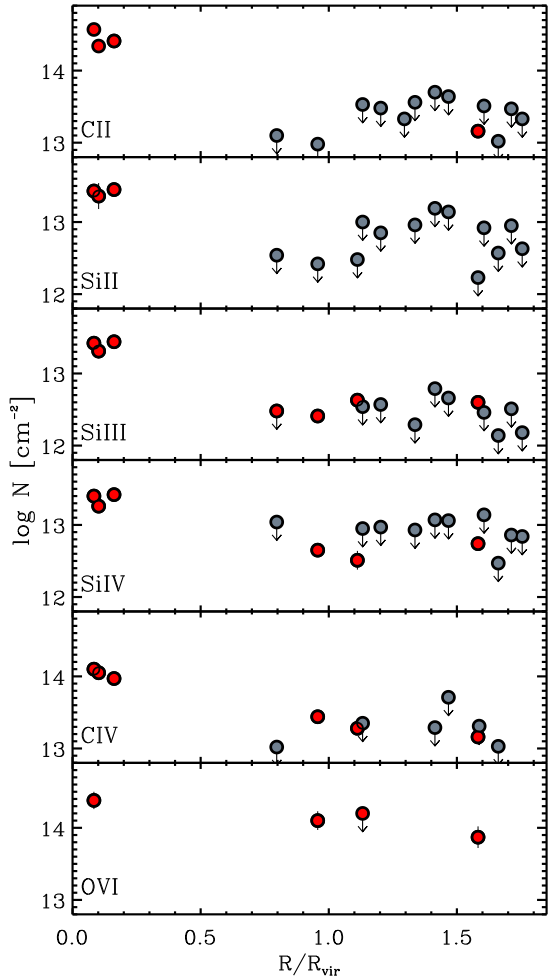}  
  \caption{Total column density of various ions for the M31 CGM component along each sightline as a function of the impact parameter.  Data with no visible error bars mean the errors are smaller than the circles. Gray data with downward arrows are $3\sigma$ upper limits. Red data with downward arrows indicate that the absorption is detected in a single transition of a single ion. 
 \label{f-nrho}}
\end{figure}

\begin{figure}[tbp]
\epsscale{1.} 
\plotone{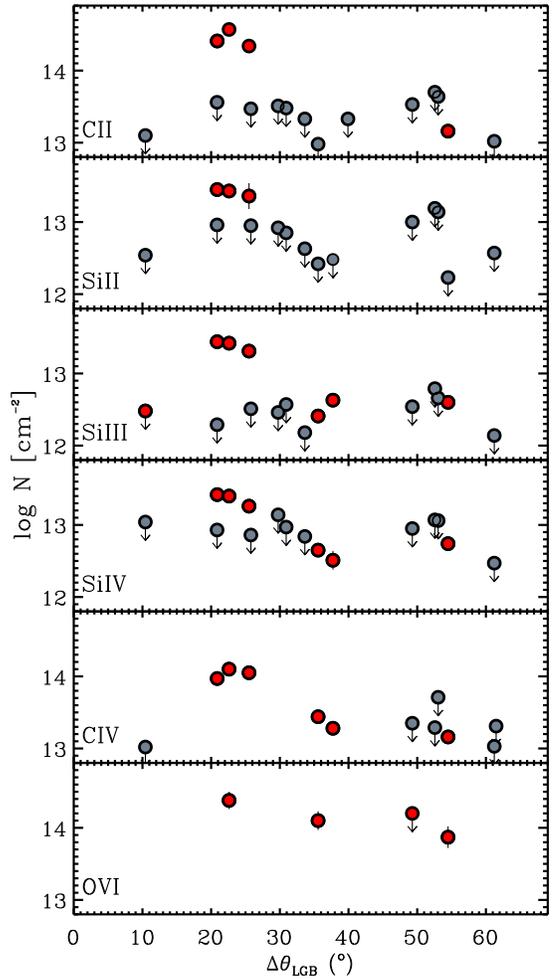}  
  \caption{Total column density of various ions for the M31 CGM component along each sightline as a function of the angular distance from the Local Group barycenter. Data with no visible error bars mean the errors are smaller than the circles.  Gray data with downward arrows are $3\sigma$ upper limits. Red data with downward arrows indicate that the absorption is detected in a single transition of a single ion. 
 \label{f-nrholg}}
\end{figure}

In Fig.~\ref{f-nrho}, we show the total column densities of the components at $-300 \la v_{\rm LSR}\la -150$ \km\ for \cii, \siii, \siiii, \siiv, \civ, and \ovi\ as a function of the projected distances from M31, $R/R_{\rm vir}$ (using the values listed in Table~\ref{t-ions}). For all the species, we observe a decrease in the column densities as $R$ increases. We stress that this relationship between $N$ and $R$ seen in all the ions is not trivial and is extremely unlikely to happen by chance. In Fig.~\ref{f-nrholg}, we show the column densities for the same component but now as a function of the angular separation from the LG barycenter ($\Delta \theta_{\rm LGB}$) \citep[where for the LG barycenter, we use $l=147\degr$ and $b=-25\degr$ from][]{einasto82,blitz99}. There is no relationship between $N$ and $\Delta \theta_{\rm LGB}$. Although we do not show it here, there is no trend of $N$ with $R$ in \siiii\ and \civ\ for the component at  $-430 \la v_{\rm LSR}\la -300$ \km\ (i.e., the MS component).

Using \siiii\ as a sensitive probe of the diffuse ionized gas \citep[e.g.,][]{collins09,shull09,lehner12}, Fig.~\ref{f-nrho} also shows that \siiii\ at $-300 \la v_{\rm LSR}\la -150$ \km\ is detected toward all the sightlines within about $R_{\rm vir}$, while only 2 of 11 sightlines beyond $R_{\rm vir}$ have a detection of \siiii. As we argue in \S\ref{s-velm31}, these two sightlines could be contaminated by the MS. This abrupt change in the detection rate of M31 near $R_{\rm vir}$ is again a non-trivial property. 

The simplest interpretation of the decrease in the column densities and drop in the detection rate with $R$ is that the absorption seen at $-300 \la v_{\rm LSR}\la -150$ \km\  traces the CGM of M31. 

\subsubsection{\hi\ Emission and UV Absorption} \label{s-hicgm}
The CGM of M31 was previously observed in \hi\ 21-cm emission \citep{thilker04,braun04,westmeier05,westmeier08,lockman12,wolfe13}. Several studies have also reported on the \hi\ observations around M33 that occupies the CGM of M31 \citep{grossi08,putman09}. The M31 observations found a concentration of \hi\ clouds within 50 kpc of M31 at $-520 \le v_{\rm LSR} \la -150$ \km. For some of these clouds, there is evidence for interaction with M31 or M33 \citep{thilker04,putman09}. Both the spatial and kinematic properties of the HVC population around M31 strongly suggest its association with M31 \citep[e.g.,][]{thilker04,westmeier08}. 

From the tabulated velocity components of the \hi\ emission associated with the CGM of M31 by \citet{westmeier08} ($R<0.2 R_{\rm vir}$), \citet{lockman12}, and \citet{wolfe13} ($R\ga 0.2 R_{\rm vir}$), we estimate $v_{\rm M31}$ and plot them in Fig.~\ref{f-vm31} with upside-down triangle symbols. It is apparent that there is also an overlap in the velocities seen in \hi\ emission and UV absorption. While none of our sightlines passes within the \hi\ cloud contours, some of the \hi\ clouds within $R<0.2 R_{\rm vir}$ from M31 and associated with M31 have emission at $-300  \la v_{\rm LSR} \la -150$ \km, i.e., at velocities seen in the UV absorption (see Fig.~\ref{f-vm31}). Although some \hi\ has been observed at LSR velocities more negative than $-450$ \km\ ($v_{\rm M31} \simeq -200$ \km) near M31 (mostly near the southern edge of M31 where these velocities are expected from the M31 co-rotating disk gas observations by \citealt{chemin09}), we did not find these in the current sample, neither in the inner ($R\la0.2 R_{vir}$) nor in the distant regions. \citet{rao13} found absorption at these velocities along their sightlines near the M31 disk using low resolution COS G140L data, but all their detections are either within or near the disk of M31, all along its major axis (see Fig.~\ref{f-map}). We finally note that all but one of the M31 dwarf satellites have also LSR velocities that are not more negative than $-450$ \km\ \citep[see Table~2 in][]{mcconnachie12}.

\subsection{Other Origins?}

We have presented several properties of the absorption at $-300 \la v_{\rm LSR}\la -150$ \km\ that favor an origin in the CGM of M31. As for the \hi\ association with the CGM of M31, we do not rely on a single property to relate this gas  to the M31 CGM. It is the combination of the properties laid out in the previous sections  that strongly suggests that the gas at $-300 \la v_{\rm LSR}\la -150$ \km\ is associated with M31. 

We have ruled out that the MS could be the origin of gas at these velocities in the directions of the sightlines in our selected sample. Neither the MW CGM (beyond 20 kpc) nor the LG appears to be a good candidate for the origin of the absorption at  $-300 \la v_{\rm LSR}\la -150$ \km. As shown in the observational cross-correlation of absorbers and galaxies at $z\la 0.5$ \citep[e.g.,][]{lanzetta95,chen01,wakker09,prochaska11,werk14} and in simulations \citep[e.g.,][]{ford14}, the column density of metal lines is expected to drop as a function of the impact parameter from the galaxies and the covering fraction is much smaller at $R>R_{\rm vir}$ than at $R<R_{\rm vir}$. These properties are observed for the gas at $-300 \la v_{\rm LSR}\la -150$ \km\ if it is associated with M31. No trend is observed with the angular separation from the barycenter of the LG (none is really expected, but none is observed by chance either). If we consider any other observed components (MS, MW disk, MW HVCs) and plot the column density of each of these components as a function of the projected distance from M31, no relationship or trend is observed. Finally, the similarity of the velocities for the galaxies associated with M31 and the absorption at  $-300 \la v_{\rm LSR}\la -150$ \km\ is remarkable. We therefore conclude that the most likely origin of the absorption at  $-300 \la v_{\rm LSR}\la -150$ \km\ is the CGM of M31 and we now proceed to determine in detail its properties. 

\section{The Properties of the CGM of M31}\label{s-prop}
In this section, we derive gas properties (kinematics, ionization, mass) of the uncontaminated component probing the M31 CGM at $-300 \la v_{\rm LSR}\la -150$ \km\ (or $-7 \le v_{\rm M31} \le +110$ \km\ in the M31-centric velocity frame, see Fig~\ref{f-vm31}). As presented in \S\ref{s-velm31}, it is plausible that the M31 CGM gas is also present at $-450 \la v_{\rm LSR}\la -300$ \km\ ($-122 \le v_{\rm M31} \le -34$ \km), but the absorption in this component is very likely largely contaminated by the MS. Hence, there is likely presence of CGM gas at both positive and negative velocities in the M31-centric velocity frames, but we can only determine the properties of the absorption at positive velocities.

\subsection{Kinematics}\label{s-kin}

Within the uncertainties, the absorption observed in the CGM gas at $<0.2 R_{\rm vir}$ is at M31-centric velocities $+50 \la v_{\rm M31} \la +100$ \km. There is an apparent trend that $v_{\rm M31} $ decreases at larger $R$ where at $R\ge 0.8 R_{\rm vir}$, $0 \la v_{\rm M31} \la +30$ \km.  The velocities toward the 7 targets within $1.1 R_{\rm vir}$ are within the expected velocities from an extension of a flat rotation curve for M31 to about its virial radius \citep[e.g.,][]{chemin09}. The remaining target, NGC7469 at  $1.6 R_{\rm vir}$ is at an azimuthal angle similar to some of the other $R_{\rm vir}$ targets, and shows absorption at velocities similar to those seen toward these  targets. However, as noted in \S\ref{s-assoc}, this sightline is the most likely to be contaminated by the MS. 

Within the calibration uncertainty of COS ($\sim 10$--$15$ \km), the high and low ions are observed at similar velocities along each sightline. The CGM gas is thus multiphase as observed in the CGM of other galaxies \citep[e.g.,][]{werk14,bordoloi14}. 

In Fig.~\ref{f-vm31}, we also show the expected escape velocity as a function of $R$ for a $10^{12}$ M$_\sun$ point mass (the dark matter halo of M31 is $(1.4 \pm 0.4) \times 10^{12}$ M$_\sun$, e.g., \citealt{marel12}).  All the detected CGM gas is found at velocities below the escape velocity even at large $R$. This holds even if we increase $v_{\rm M31}$ by $\sqrt{3}$ to account for unconstrained projection effects. \citet{mcconnachie12} demonstrated that it also holds for most of the galaxies using their 3D distances (instead of the projected distances used here). Therefore, the CGM gas probed in our sample at both small and large $R$ is gravitationally bound to M31.

\subsection{Ionization}\label{s-ion}
For the sightlines within  $R<0.2 R_{\rm vir}$, there is no strong evidence for significant variation in the ionization levels or gas surface densities since the column densities are similar within 1$\sigma$ for each studied ion (see Table~\ref{t-ions} and Fig.~\ref{f-nrho}). However, as shown in Fig.~\ref{f-nrho} and already noted in \S\ref{s-assoc}, the column densities of the M31 CGM component decrease as $R$ increases for all the ions beyond $0.2 R_{\rm vir}$. The gas also becomes more highly ionized as $R$ increases, as the singly-ionized species are mostly not observed beyond $0.2 R_{\rm vir}$ (although note the lack of data between  $0.2$ and  $0.8 R_{\rm vir}$), while the higher ions (\siiii, \siiv, \civ, \ovi) are detected to $\sim R_{\rm vir}$ and possibly at $1.6 R_{\rm vir}$. The drop in column density between $0.2 R_{\rm vir}$ and $\ge 0.8 R_{\rm vir}$ is a factor $\sim 10$ for \siiii, at least a factor 10 for the singly ionized species, a factor $\sim 5$ for \siiv\ and \civ, and a factor $\sim 2$ for \ovi. There is no detection of singly ionized species beyond $0.2 R_{\rm vir}$, except possibly at $1.6 R_{\rm vir}$ toward NGC7469 (which is possibly some MS gas, see \S\ref{s-assoc}). 

The gas is therefore more highly ionized at larger $R$ than in the inner region. This can be directly quantified using the ratios of several ions of the same element. At $R < 0.2 R_{\rm vir}$, we find $N_{\rm CII}/N_{\rm CIV}\,\ga 2$--3 and $N_{\rm SiII} \sim N_{\rm SiIII} \sim N_{\rm SiIV}$; however, at $R>0.8 R_{\rm vir} $, we estimate $N_{\rm CII}/N_{\rm CIV}\,\la 1$ and $N_{\rm SiII}/ N_{\rm SiIV} < 0.6$. We also note that the observed ratio $N_{\rm CIV}/N_{\rm SiIV} \sim 5$ throughout the CGM of M31 is  on the high end of those observed in the MW disk and halo gas \citep[e.g.,][]{lehner11a,wakker12}. 

We can directly limit the fractional ionization of the gas by comparing the \oi\ column to the column densities of \siii, \siiii, and \siiv\ \citep{lehner01,zech08}. \oi\ is an excellent proxy of \hi\ since they have nearly identical ionization potentials and are strongly coupled through charge exchange reactions \citep{field71}. O and Si have also the same nucleosynthetic origin (they are both $\alpha$-elements) and have similar levels of dust depletion in the diffuse gas \citep{savage96,jenkins09}. Whereas \oi\ arises only in neutral gas, \siii\ is found in both neutral and ionized gas and \siiii\ and \siiv\ arise only in ionized gas. Therefore a subsolar ratio of  [\oi/Si$] = \log (N_{\rm OI}/N_{\rm Si}) - \log ({\rm O/Si})_\sun $ is expected if the ionization is important.\footnote{We also note that in regions with significant dust, the gas-phase O/Si is expected to be supersolar.} Except for HS0033+4300 and UGC12163, the S/N of the data allows us to place stringent limits on the column density of \oi.

 Using the results listed in Table~\ref{t-ions} we find: [\oi/Si$]  <-1.4$ and $<-1.1$ toward RXJ0048.3+3941 and HS0058+4213 at $R<0.2R_{\rm vir}$, [\oi/\siii$]<-0.1$ toward 3C66A at $0.8 R_{\rm vir}$, [\oi/Si$]<-0.6$ toward MRK335  and  $<-0.5$  toward PG0003+158 at about $R_{\rm vir}$, and [\oi/Si$]<-0.9$ toward NGC7469  at $1.6 R_{\rm vir}$. This implies hydrogen ionization fractions $\ge 93$--$97\%$ at $R<0.2R_{\rm vir}$.  It is very likely that the large impact parameter regions are almost entirely ionized too, especially since the mixture of gas-phases favors the more highly ionized phases. However, the weaker absorption at larger $R$ produces less stringent limits on the ionization fraction. The CGM gas of M31 is therefore mostly ionized toward all the targets in our sample. It is also more highly ionized at larger $R$. 

The detections of \civ\ and \ovi\ are also particularly important since they provide  some evidence for $\sim 10^5$ K gas. It is also an indirect evidence for an extended corona around M31 if thermal instabilities in the hot corona are important or the high ions are at the interface between cool and hot gas. For example, thermal instabilities in a hot corona can produce a warm multiphase, extended CGM if the thermal instability growth time is a factor $\la 10$ times smaller than the free-fall time of the hot gas \citep{mccourt12,sharma12}.  We also note there is detection of \ovi\ ($\log N_{\rm OVI} \simeq 13.9$--14.1) toward  MRK352, PG0052+251, and MRK357 at LSR velocities of about $-190$ \km\ \citep[see the component marked ``LG'' in Table~1 of][]{sembach03}, possibly consistent with an origin from the M31 CGM. These targets are at $14\degr \la {\rm RA} \la 21\degr$, $23\degr \la {\rm DEC} \la 32\degr$, and probe the M31 CGM at $0.4 < R/R_{\rm vir}<0.9$. They are not in our sample because only \fuse\ data exist (see \S\ref{s-sample}). Future COS observations of these targets would be extremely valuable to confirm such an interpretation.   The hot corona also provides an alternative explanation for the existence of \hi\ clouds in the CGM of M31 to that proposed by \citet{wolfe13} where they argued that the \hi\ clouds are embedded in an intergalactic filament.

\subsection{Covering fractions}\label{s-cover}
Within $R_{\rm vir}$, the covering fraction for \siiii\ and the high ions is close to unity. Each sightline shows \siiii\ absorption from the M31 CGM gas, and hence the covering fraction is 100\% ($>70\%$ at the 68\% confidence level for a sample size of 5) for a $3\sigma$ sensitivity limit of $\log N_{\rm SiIII} \simeq 12.4$ (assuming FHWM\,$\simeq 40$ \km). For \civ, 4/5 of the sightlines show absorption down to a sensitivity of $\log N_{\rm CIV} \simeq 12.9$, which corresponds to a covering fraction of 80\% ($>48\%$ at the 68\% confidence level for a sample size of 5). For \ovi, we have only two targets in the inner-most and outer-most regions, and they all show \ovi\  absorption associated with the M31 CGM gas. 

Beyond $R_{\rm vir}$, the covering fraction for \siiii\ and the high ions is much smaller in view of the many non-detections. If we a set a $3\sigma$ sensitivity limit of $\log N_{\rm SiIII} \simeq 12.6$ based on the detection of \siiii, then the detection is rate is $10$--$20\%$ for $1\la R/R_{\rm vir} \la 1.8$ ($<35\%$ at the 68\% confidence level for a sample size of 10). For \civ\ with a sensitivity limit of $\log N_{\rm CIV} \simeq 13.3$, the covering fraction is about $20$--$30\%$  ($<60\%$ at the 68\% confidence level for a sample size of 6). 

For the singly ionized species, the covering fraction is  about 100\% for $R \le 0.2 R_{\rm vir}$ and $\ll 1$ at $R \ga 0.8 R_{\rm vir}$ (see Fig.~\ref{f-nrho}). It would be critical to gain information between $0.2 R_{\rm vir}$ and $0.8 R_{\rm vir}$ to determine at which projected distance from M31 the covering fraction of singly ionized species becomes much smaller than 1. 

\subsection{Metal and Gas Mass}\label{s-metal}

The metal mass of the CGM is independent of the metallicity and can be directly determined from the column densities estimated in Table~\ref{t-ions}. Following \citet{peeples14}, the mass of metals in the CGM is $M_{\rm Z} = \int 2\pi R \Sigma_{\rm Z}(R) dR$. Using Si as a tracer of the metals,  the mass surface density of metals is defined as $\Sigma^{\rm Si}_{\rm Z} = \mu_{\rm Si}^{-1}\, m_{\rm Si} N_{\rm Si} $ where  $\mu_{\rm Si}=0.064$ is the solar mass fraction of metals in silicon \citep[i.e., $12+\log ({\rm Si/H})_\sun = 7.51$ and $Z_\sun = 0.0142$ from][]{asplund09}, $m_{\rm Si} = 28 m_{\rm p}$, and $N_{\rm Si} =  N_{\rm SiII}  +  N_{\rm SiIII}  + N_{\rm SiIV} $. We use Si here because we have information on its 3 dominant ionization stages in the $T<7\times 10^4$ K ionized gas, requiring no assumptions on the ionization of the cool gas.  

We first consider the region $R<0.2 R_{\rm vir}$ where we noted in \S\ref{s-ion} that the column densities for each ion toward the 3 sightlines in this region do not vary much (see Fig.~\ref{f-nrho}). Using the column densities summarized in Table~\ref{t-ions} and averaging the values for \siii, \siiii, and \siiv\ for sightlines at $R<0.2 R_{\rm vir}$, we find $\langle N_{\rm Si}\rangle  = 7.4\times 10^{13}$ cm$^{-2}$. The mass surface density of metals in the CGM of M31 is then:
$$
\log \frac{\Sigma^{\rm Si}_{\rm Z}}{\rm M_\sun kpc^{-2}} = 2.6,
$$
which is similar to the observed range of metal surface densities found for $z\sim 0.2$ $L^*$ galaxies \citep[$\log \Sigma_{\rm Z} =3.2$--$2.0$ at $R=26$--$50 $ kpc][]{peeples14}. Assuming a covering fraction of 100\% based on the detections of \siii, \siiii, \siiv\ in the spectra of the 3 targets in our sample, the CGM metal mass of M31 within $R\le 0.2 R_{\rm vir}$ ($R\le 50$ kpc) is 
$$
M^{\rm Si}_{\rm Z} = 2.1 \times 10^6  \,\,{\rm M}_\sun.
$$
This is a factor $\sim$25 smaller times than the dust mass in the disk of M31, $M_{\rm d}({\rm disk}) = 5.4 \times 10^7$ M$_\sun$ \citep{draine14}. The corresponding total gas mass of the CGM of M31 within 50 kpc can be estimated from the metal mass as: 
$$
M_{\rm g} = 1435 M_{\rm Si} \Big(\frac{Z_\sun}{Z}\Big) = 2.9 \times 10^9\, \Big(\frac{Z_\sun}{Z}\Big)\,\,{\rm M}_\sun.
$$ 
Although we do not know the metallicity $Z/Z_\sun$ of the M31 CGM gas, it is unlikely to be much larger than solar. For the dense region of the CGM of galaxies at $z\la 1$, \citet{lehner13} found that metallicities range between about $\la 0.01  Z_\sun$ and $\sim 3 Z_\sun$, the highest metallicities being associated only with massive outflows \citep{tripp11}. In the disk of M31, \citet{sanders12} found a scatter between $0.3 \la Z/Z_\sun \la 4$. The highest allowed metallicity values could therefore not decrease the mass by more than a factor $\sim 3$--$4$.  On the other hand, the total mass could be much larger if the metallicity is sub-solar. Within $0.2 R_{\rm vir}$, the gas mass in the CGM of M31 is therefore comparable to that in its disk, since $M_{g}({\rm disk}) \approx 6 \times 10^9$ M$_\sun$ using the recent results from \citet{draine14}. 

This mass does not include the more highly ionized gas  traced by \ovi\ and \civ. If we use \civ\ \citep[where $\mu_{\rm C}=0.23$,][]{asplund09} and assume an ionization fraction of $f_{\rm CIV} = 0.3$ \citep[which is an upper limit on $f_{\rm CIV}$ in collisionally or photoionized gas, see e.g.,][]{gnat07,oppenheimer13}, we find $M^{\rm CIV}_{\rm Z} \ge  1.4 \times 10^6 (0.3/f_{\rm CIV}) (R/50\,{\rm kpc})^2$\,M$_\sun$, i.e., the highly ionized CGM gas has a similar mass as the cool photoionized CGM gas. 

\begin{figure}
\epsscale{1.2} 
\plotone{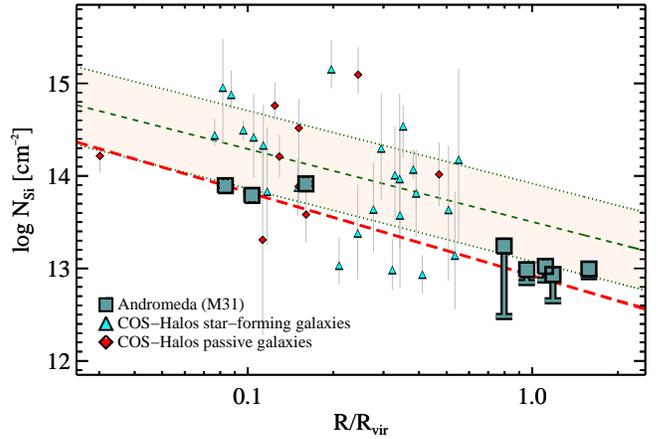}
  \caption{Comparison of the total column density of Si versus the impact parameters from M31 and the COS-Halos galaxies from \citet{werk14}. For the COS-Halos data, the vertical error bar shows the range allowed by the Cloudy photoionization model \citep[see][]{werk14}. For M31, the values of $N_{\rm Si}$ are directly constrained by our estimates on  $N_{\rm SiII}$,  $N_{\rm SiIII}$, and  $N_{\rm SiIV}$ from the observations; the error bars are less than the size of the square and the vertical blue bars at large impact parameters include the range of possible values if the non-detections are either near their $3\sigma$ limits or so low as to be negligible. The power-law fit from a linear regression analysis to the data is shown as a green dashed line for the COS-Halos sample  (with the light beige shaded area representing the $1\sigma$ uncertainty in the fit parameters) and a  red dashed line for the M31 sample. 
 \label{f-werk}}
\end{figure}

The current sample of sightlines is too small to determine how the mass changes with radius or azimuth angle. There is also some uncertainty at $R\ga R_{\rm vir}$ owing to a plausible contamination from the MS (see \S\ref{s-assoc}). Nevertheless, there is undoubtedly  a sharp  decrease in the column densities at  $R \ga R_{\rm vir}$ compared to $R<0.2 R_{\rm vir}$, and this drop in $N$ would be even larger if there is some contamination from the MS. This is displayed in Fig.~\ref{f-nrho} for all the ions and in Fig.~\ref{f-werk}, where we show the total column density of Si (where Si\,$=$\,\siii+\siiii+\siiv) as a function of $R/R_{\rm vir}$. For the data at $R\ge 0.8 R_{\rm vir}$, the vertical bars include the range of possible values if the non-detections are either near their $3\sigma$ limits or so low as to be negligible (see Table~\ref{t-ions} and figures in the Appendix). With no information between the inner and outer regions of the CGM, a linear fit provides the simplest model for characterizing this observed change in the column density of Si with $R$, which yields $N_{\rm Si}(R) = 10^{12.9}\,  (R/R_{\rm vir})^{-0.9}$ cm$^{-2}$.  This fit is of course poorly constrained and would require more data to model accurately $N(R)$ and hence $M(R)$. In Fig.~\ref{f-werk}, we also overlay the COS-Halos results from \citet{werk14}. The experiments that led to these results are very different since for COS-Halos, there is one data point per galaxy while there are 7 measurements (and more including the non-detections) for M31. Even given this difference, similar trends are observed. The column densities for M31 are systematically on the low side of the COS-Halos distribution, possibly because some of the absorption may be blended, in particular,  with the MS absorption, and hence not taken into account in the total column density (see \S\ref{s-velm31}). For example, making the unreasonable assumption that all the observed absorption arises from the M31 CGM toward HS0058+4213 at $R\sim 0.2 R_{\rm vir}$, we find $\log N_{\rm Si}>14.8$. It is quite plausible that the total column density of Si for the M31 CGM lies between that limit and 13.9 dex, a range of values consistent with COS-Halos. 

To estimate the mass within the volume at $R_{\rm vir}$, we use the result from our fit to integrate $M(R)$ to $R_{\rm vir}$. We find $M^{\rm Si}_{\rm Z}(R_{\rm vir}) \sim 1.4 \times 10^7$ M$_{\sun}$ and $M_{\rm g}(R_{\rm vir}) \sim 2.1 \times 10^{10}$  M$_{\sun}$. As we saw above, the mass of the highly ionized gas is about of the same order. It is also likely that this represents only a lower limit to the total mass of the M31 CGM gas since some of the absorption is likely contaminated by the Magellanic Stream (see \S\ref{s-velm31}).  As the stellar mass of M31 is  $10^{11}$ M$_{\sun}$ \citep[e.g.,][]{geehan06,tamm12},  the  mass of the diffuse weakly and highly ionized CGM of M31 is therefore a substantial fraction of the total baryonic mass of M31. 

\section{Discussion}\label{s-disc}

Prior to this work, the CGM of M31 beyond its optical radius has been only explored with deep \hi\ 21-cm emission observations at a sensitivity of $\log N_{\rm HI} \ge 17$ \citep{thilker04,braun04,lockman12,wolfe13}. Within about 50 kpc ($0.2 R_{\rm vir}$), \citet{thilker04} found low $N_{\rm HI}$ filamentary  gas within $\sim 80$ \km\ of M31 systemic velocity, i.e., with  a velocity separation similar to that seen in the UV data presented here (see Fig~\ref{f-vm31}). They derived a total mass for the \hi\ gas observed through the GBT $9\farcm1$ beam of $3$--$4 \times 10^7$\,M$_\sun$. At $R>50$ kpc, the $49\arcmin$ Westerbork \hi\ survey by \citet{braun04} suggested a filament of \hi\ gas between M31 and M33, but this structure has dissolved into very small clouds of diameters less than a few kpc and masses of a few $10^5$ M$_\sun$ when observed with the GBT \citep{lockman12,wolfe13}. While the \hi\ mass is still poorly constrained, it is evident the mass of the M31 CGM is dominated by the diffuse ionized gas (see \S\ref{s-metal}). Our sample is still small within $R_{\rm vir}$, but each sightline with its pencil beam shows detection of M31 CGM gas for at least \siiii, with most showing absorption from multiple ions (\siii, \siiii, \siiv, \cii, \civ, \ovi). It would be of course important to fill the radial and azimuthal spaces with more COS observations of QSOs behind the CGM of M31 in order to confirm these results and more accurately characterize its physical structure. Nevertheless the present observations provide already strong evidence that the CGM of M31 is filled with multiphase, diffuse ionized gas (see \S\ref{s-cover}).  We do not detect any \oi\ absorption in any of the QSO spectra piercing the CGM of M31, which puts stringent limits on the ionization level, \hii/\hi\,$\ga 93$--$97\%$ (see \S\ref{s-ion}), consistent with the small covering fraction of \hi\ detectable with 21-cm emission observations. Although deep \hi\ emission observations with $\log N_{\rm HI} >17$ reveal only the tip of the iceberg of the CGM, it will be critical that future radio \hi\ surveys can achieve this type of sensitivity with a good angular resolution in order to bring to light the spatial distribution of the \hi\ gas beyond the optical radii of galaxies. The present COS UV sample provides therefore the first strong evidence in the LG for CGM gas beyond 50 kpc  \citep[see][]{lehner12}. 

As displayed in Fig.~\ref{f-vlms}, the associated components with the CGM of M31 are found at $-300 \la v_{\rm LSR} \la -150$ \km\ ($-7 \le v_{\rm M31} \le +110$ \km). As shown in Fig.~\ref{f-vm31}, the comparison with the velocity distribution of the dwarf satellites suggests that some of the absorption observed at $-121 \le v_{\rm M31} \le -34$ \km\ ($-450 \la v_{\rm LSR} \la -300$ \km) is a mixture of the MS and M31 CGM components, where the absorption is dominated by the MS (see \S\ref{s-velm31}). We also find that $\sqrt{3} v_{\rm M31}< v_{\rm esc} $ (see Fig.~\ref{f-vm31} and \S\ref{s-kin}), implying that the gas is bound to M31 even at large $R$. 

These results could suggest that the CGM of the MW might be similarly large, but to characterize it will be difficult, since in view of our findings for the M31 CGM, the MW CGM absorption may also be dominated by low-velocity halo clouds (LVHCs, $|v_{\rm LSR}| \la 90$ \km), where the absorption is strongly blended with the MW disk and low halo. Except for the MS, most of the HVCs and iHVCs have been indeed found to be within 5--15 kpc (see, e.g., \citealt{thom08,wakker08,lehner11,lehner12}, and also \citealt{richter12} who found a characteristic radial extent of 50 kpc for the \hi\ HVCs of the MW and M31 using a model with a radial exponential decline of the mean \hi\ volume-filling factor).  The LVHCs may be the best candidate for an extended MW CGM \citep{peek09}, along with some of the very high-velocity clouds not associated with the MS \citep{lehner12}. 

We determine that the baryonic mass of the weakly and highly ionized CGM gas of M31 is at least about 30\% of the total baryonic mass of M31 (see \S\ref{s-metal}), but this does not include the hot $\ge 10^6$ K CGM coronal component.  The ubiquitous detection of high ions in our sample suggests the presence of hot ($>10^6$ K) diffuse gas surrounding M31 within its virial radius and possibly beyond (see \S\ref{s-ion}) if the production mechanisms of the high ions are dominated by thermal instabilities in the hot corona or interfaces between the cool (\siii, \siiii) and putative hot gas (see \S\ref{s-ion}). In cosmological simulations, substantial amounts of \ovi\  are produced through collisional ionization in the CGM of galaxies as it transistions from cooling of hot gas \citep{oppenheimer12,cen13,ford13,ford14}, although some of the \ovi\ could be photoionized in very low densities at impact parameters $\ge 100$ kpc ($\ge 1/3 R_{\rm vir}$) \citep{ford14}. \citet{cen13} show that collisional ionization dominates the production of strong ($N_{\rm OVI}> 10^{14}$ \cmm) \ovi\ absorbers. For M31 we only find $N_{\rm OVI} <10^{14} $\ \cmm\ beyond $R_{\rm vir}$ (see Fig.~\ref{f-nrho} and \S\ref{s-ion}). 

The hot galaxy coronae are one of the fundamental predictions of galaxy formation models \citep{white78,white91}, but their direct detection has been very difficult. Progress has been made recently with the detection of diffuse X-ray emission that appears to extend to about 30--70 kpc around a handful of massive, non-starbursting galaxies \citep{anderson11, bodgan13a,bodgan13b} or in stacked images of galaxies \citep{anderson13}. The mass estimate for these hot halos at these radii are about a few times $10^9$ M$_\sun$, which is comparable to the mass found in the cooler ($< 10^5$ K) gas of the CGM of M31 (see \S\ref{s-metal}), and hence the total mass of the CGM of M31 including all the gas-phases could be as large as $\sim 10^{10}$ M$_{\sun}$ within 50 kpc. Beyond $50$ kpc, the CGM is too diffuse to be traced with X-ray imaging, even though a large mass could be present.  Extrapolating to about the virial radius, \citet{anderson11} estimate that the hot halo mass of the massive spiral galaxy NGC1961 might be about $10^{11}$ M$_{\sun}$ (the stellar mass of NGC1961 is $\sim 3$ times that of M31), a factor 5 larger than the mass of the cool CGM of M31 for the volume within $R_{\rm vir}$ (see \S\ref{s-metal}). For the MW, \citet{gupta12} argue that the mass of the $2\times 10^6$ K CGM out to 160 kpc could be as high as  $10^{11}$ M$_{\sun}$. However, there is still a large disagreement on the interpretation of the X-ray observations and their implications for an extended hot  MW CGM \citep{gupta12,gupta14,wang12,henley14,sakai14}. The most recent estimate for the MW implies a smaller mass for the $2\times 10^6$ K MW CGM gas of about  $4\times 10^{10}$ M$_{\sun}$ within 50 or 240 kpc \citep{miller15}. Based on these X-ray observations, a substantial mass of the M31 CGM could also be present in its hot ($>10^6$ K) diffuse corona.

\begin{figure}
\epsscale{1.2} 
\plotone{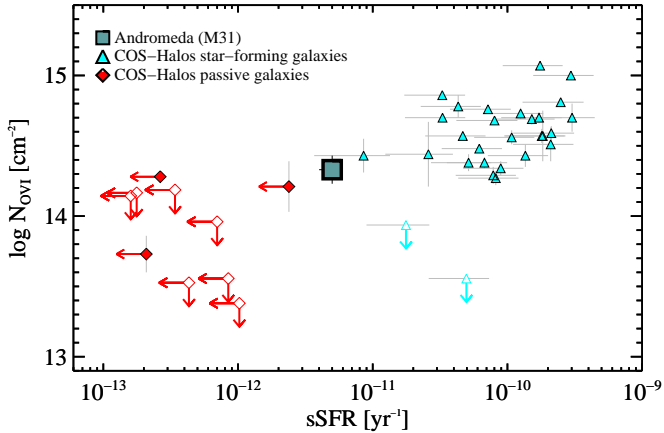}  
  \caption{Comparison of the total column density of \ovit\ versus the specific star-formate rate of the galaxy between M31 and the COS-Halos program \citep{tumlinson11}. For M31, we only consider the sightline with $R =25$ kpc since the COS-Halos targeted galaxies with $R\le 150$ kpc.  \label{f-ovi}}
\end{figure}

Our results echo the findings from the COS-Halos survey of $L*$ galaxies at $z\sim 0.2$ \citep{tumlinson11,werk13,werk14,peeples14}. In Figs.~\ref{f-werk} and \ref{f-ovi}, we reproduce two COS-Halos figures with our  M31 results included. Fig.~\ref{f-werk} was already presented in \S\ref{s-metal}, which shows a similar distribution for $N_{\rm Si}$ as a function of $R/R_{\rm vir}$ between the COS-halos and the present study, which implies similar masses for their cool (probed by \siii, \siiii, \siiv) and warm-hot (probed by \civ\ and \ovi) ionized CGM. The typical dark matter halos of the COS-Halos galaxies is $10^{12}$ M$_\sun$, similar to M31 \citep[e.g.,][]{marel12}. In Fig.~\ref{f-ovi}, we show $N_{\rm OVI}$ against the specific star-formation rate of the COS-Halos galaxies and M31.\footnote{As for Si, the total amount of \ovit\ could be larger if some of the \ovit\ absorption of the CGM of M31 is blended with the MS or the MW. However, for \ovit, the total column would only increase by about 0.3 dex.} M31 is right  between the COS-Halos passively-evolving and star-forming galaxies where star-forming systems have typically $\mlnovi \ga 14.3$. The column density of \ovi\ through the M31 CGM is therefore  consistent with the transition observed between the COS-Halos passively-evolving and star-forming galaxies, providing additional evidence -- independent of that based on its colors -- that M31 might be indeed transitioning from a blue to a red galaxy \citep{mutch11,dalcanton12}. We emphasize that although the mass and luminosity of M31 are comparable to the COS-Halos galaxies, their environments might be quite different: M31 lives in a galaxy environment with close by  companions (M32, M33, MW, etc.) while the COS-halos galaxies were selected to be fairly isolated. The influence of the MW (750 kpc from M31) on the CGM within 300 kpc of M31 may be thus be relatively small.  While the M31 CGM has properties similar to other $L*$ galaxies,  we find CGM masses at least about 5--6 times larger than estimated for the sub-$L*$ galaxies at $z \sim 0$ \citep{bordoloi14}.

The origin of this diffuse ionized CGM gas is an open question. The distribution and composition of the  CGM gas have been addressed in several  high-resolution cosmological hydrodynamic simulations with various level of star formation and feedback prescriptions \citep[e.g.,][]{klypin02,joung12,cen13,nuza14,ford14}. Notably, despite the various treatments and different levels of the feedback in these simulations, there is some  agreement regarding the distribution of the baryons in the CGM  between these simulations and  with our empirical results as we now outline. 

One of the basic requirement of all these simulations is that a large fraction  of the baryons are in the CGM of galaxies.  For example, the $\Lambda$CDM-models of the Milky and M31 by \citet{klypin02} requires that 1/4 to 1/2 of the baryons within the virial radius of the halo must not be in the disk or bulge of the MW and M31 in the absence of any feedback, which is consistent with our empirical results (see \S\ref{s-metal}).  \citet{cen13} predicts that over half of the gas mass within 150 kpc of red or blue galaxies is in the cool phase of the CGM. \citet{nuza14} studied the gas properties in the CGM of MW and M31 analogs using a cosmological simulation of the LG. They find masses for the cool CGM within $0.2 R_{\rm vir}$ and $R_{\rm vir}$ that are consistent with our results. The covering fraction of the CGM  gas with $\log N_{\rm HI}>15$ (i.e., tracing some of the  ionized gas reported in our work) in their simulation appears also to be broadly consistent with our observations, being close to 1 at $R\la 0.4R_{\rm vir}$ and then progressively decreasing to $\la 0.3$ at $R\ga R_{\rm vir}$. It is worth noting also that in their simulation as well as the more general simulations of galaxies at $z\sim 0$ \citep[e.g.,][]{cen13,ford14}, the cool CGM dominates the mass of the CGM within $0.2 R_{\rm vir}$  (by a factor $\sim 4$ relative to the mass of the hot gas), but there is a turnover at larger radii where at $R_{\rm vir}$ the mass of the hot gas is a factor 3 larger than the mass of the cool gas.  While we cannot observe the hot gas associated with the CGM of M31, we find that the gas becomes more ionized and more highly ionized at $R>0.8 R_{\rm vir}$ than at  $R<0.2 R_{\rm vir}$ (see \S\ref{s-ion}).  \citet{ford14} specifically investigated the distribution of similar ions studied in our work and find that the covering fractions of the low (e.g., \cii) and high (\civ, \ovi) ions are all typically high (near 100\%) at  $R<0.2 R_{\rm vir}$ and drops much more rapidly at higher impact parameter for the low ions  than the high ions,  again consistent with our empirical  results (see \S\ref{s-prop}). The similarity in the distribution of the column densities of \civ\ and \siiv\ with $R$ in the simulations of \citet{ford14} (see their Fig.~7) and our observations (see Fig.~\ref{f-nrho}) is striking, with a comparable trend  and magnitude in the variation of $N$ with $R$ (only \civ\ and \siiv\ are common in both studies). 

In the Nuza et al. and Ford et al. cosmological simulations, the CGM gas belongs to the ambient medium rather than being associated with satellites. \citet{ford14} also show most of the metal mass comes from recycled accretion at any $R$ (i.e.,  gas that was once ejected in a wind at least once before), but this is different for the baryons where if the total mass at $R<0.2 R_{\rm vir}$ is largely dominated by recycled accretion, at  $R>0.2 R_{\rm vir}$ the ambient gas (i.e., gas is  not going to accrete onto a galaxy and that have never been in a wind by $z \sim 0$) dominates the total mass. 

This comparison between cosmological models and our observations is extremely encouraging and demonstrates that a larger sample of targets observed with COS that populate in particular the $0.2 \le R/R_{\rm vir}\la 1$ region of the M31 CGM would make M31 a perfect testbed for theories of galaxy formation and evolution. With our present sample, we cannot accurately assess how the surfaces densities of the different ions change with $R$ (with an absence of data points at $0.2<R/R_{\rm vir}<0.8$) or azimuthal angle (e.g., to determine how the surface densities and kinematics vary along the major and minor projected axes of M31). This would also help us to better understand the origins of the metals and baryons in the CGM of M31.

\section{Summary}\label{s-sum}
With \hst/COS G130M and G160M and \fuse\ archival observations, we compile a sample of 18 sightlines that pierce the CGM of M31, 3 at small  ($R<0.2 R_{\rm vir}$) and 15 at large ($0.8\le R/R_{\rm vir}\la 2$) projected distances to determine the properties of the CGM at various radii. Along these sightlines, the gas seen in absorption at $-300 \la v_{\rm LSR} \la -150$ \km\ in the 3/3 inner-region sightlines and in 4/15 outer-region sightlines is neither associated with the MW (its disk or CGM) nor with the MS. These velocities are about 40--90 \km\ from the systemic velocity of M31. Only for this component do we observe a non-trivial relationship between the column densities of the observed ions (\siii, \siiii, \siiv, \cii, \civ, \ovi) and the projected distance from M31 whereby $N$ decreases with $R$. These and other arguments presented in \S\ref{s-assoc} imply that the absorption at  $-300 \la v_{\rm LSR} \la -150$ \km\ observed in the directions probed by our selected sightlines must in all likelihood arise from the CGM of the M31. We determine the following properties for the CGM of M31 traced by this absorption:

\begin{enumerate}
\item The M31 CGM gas is observed at similar velocities to that of the satellites galaxies located in the CGM of M31. These velocities are small enough relative to the escape velocity to conclude that the observed CGM gas is gravitationally bound even at large $R$. 

\item Low (\siii, \cii), intermediate (\siiii), and high ions (\siiv, \civ) are all observed in absorption toward the 3 sightlines at $R<0.2 R_{\rm vir}$, while typically only \siiii\ and high ions are observed at $R\ge 0.8 R_{\rm vir}$. The CGM gas of M31 is therefore more highly ionized at larger projected distances.

\item Within $R_{\rm vir}$, the covering fraction for \siiii\ and the high ions is close to unity. Beyond $R_{\rm vir}$, the covering fraction for \siiii\ and the high ions drops to $<20\%$. For the singly ionized species, the covering fraction is  about 100\% for $R \le R_{\rm vir}$ and $\ll 1$ at $R \ga 0.8 R_{\rm vir}$. 

\item  With sensitive limits on the column density \oi\ and its comparison to the total column density of Si (\siii+\siiii+\siiv), we show that the CGM of M31 is predominantly ionized ($\ga 95\%$).  The M31 CGM gas is therefore multiphase, dominantly ionized (i.e., \hii\,$\gg$\,\hi), and becomes more highly ionized gas at larger $R$. The presence of a large amount of multiphase gas suggests that M31 has very likely a hot corona extending all the way to $R_{\rm vir}$ and  possibly beyond.

\item We derive using \siii, \siiii, and \siiv\ a CGM metal mass of  $2.1 \times 10^6$ M$_\sun$ and gas mass of $3 \times 10^9\,(Z_\sun/Z)$ M$_\sun$ within $0.2 R_{\rm vir}$, implying a substantial mass of metal and gas within $0.2 R_{\rm vir}$. In the volume within $R_{\rm vir}$, we estimate $M_{\rm Si} \sim 1.4 \times 10^7$ M$_{\sun}$ and $M_{\rm g} \sim 2.1 \times 10^{10}$  M$_{\sun}$; there is, however, a substantial uncertainty in the masses estimated at $R_{\rm vir}$ given the lack of observations between $0.2 \le R/R_{\rm } \le 0.8$.  The highly ionized CGM gas of M31 probed by \civ\ (and \ovi) have most likely similar masses. The baryonic mass of the weakly and highly ionized CGM gas of M31  is  about 30\% of the total baryonic mass of M31 (not including the hot $\ge 10^6$ K CGM coronal gas). 

\item  The above conclusions imply that M31 has an extended, massive, multiphase CGM as observed in higher redshift $L^*$ galaxies. With the current data, there is a broad agreement between our empirical results and recent cosmological simulations in terms of baryonic mass and change in the ionization levels with projected distances. However, a larger sample will be critical to determine the properties in the $0.2\la R/ R_{\rm vir}<1$ range where currently there is no information. Despite an environment that is different from the isolated galaxies in the COS-Halos survey, the properties of the CGM of M31 is fairly typical of a $L^*$ disk galaxy that might be transitioning from a blue to red galaxy. 

\end{enumerate}

\section*{Acknowledgements}
We thank Jason Tumlinson and Jessica Werk for sharing the COS-Halos data and the IDL codes that helped producing Figs.~\ref{f-werk} and \ref{f-ovi}. We thank Kate Rubin, Mary Putman, Filippo Fraternali, Prateek Sharma, David Valls-Gabaud, and Daniel Wang for useful suggestions and discussions and the anonymous referee for providing feedback, which helped us to improve the content and clarity of our manuscript. NL and JCH acknowledge support for this research provided by  National Science Foundation under grant no. AST-1212012 and NASA through grants  HST-GO-12604, HST-AR-12854, and HST-GO-12982 from the Space Telescope Science Institute, which is operated by the Association of Universities for Research in Astronomy, Incorporated, under NASA contract NAS5-26555. BPW acknowledges support from the National Science Foundation under grant no. AST-1108913 and NASA through grant HST-GO-12604 from the Space Telescope Science Institute. All of the data presented in this paper were obtained from the Mikulski Archive for Space Telescopes (MAST). STScI is operated by the Association of Universities for Research in Astronomy, Inc., under NASA contract NAS5-26555. This research has made use of the NASA's Astrophysics Data System Abstract Service and the SIMBAD database, operated at CDS, Strasbourg, France.

\begin{appendix}
\makeatletter 
\renewcommand{\thefigure}{A\@arabic\c@figure} 

\renewcommand{\thetable}{A\@arabic\c@table} 

\begin{figure}[tbp]
\epsscale{1.0} 
\plotone{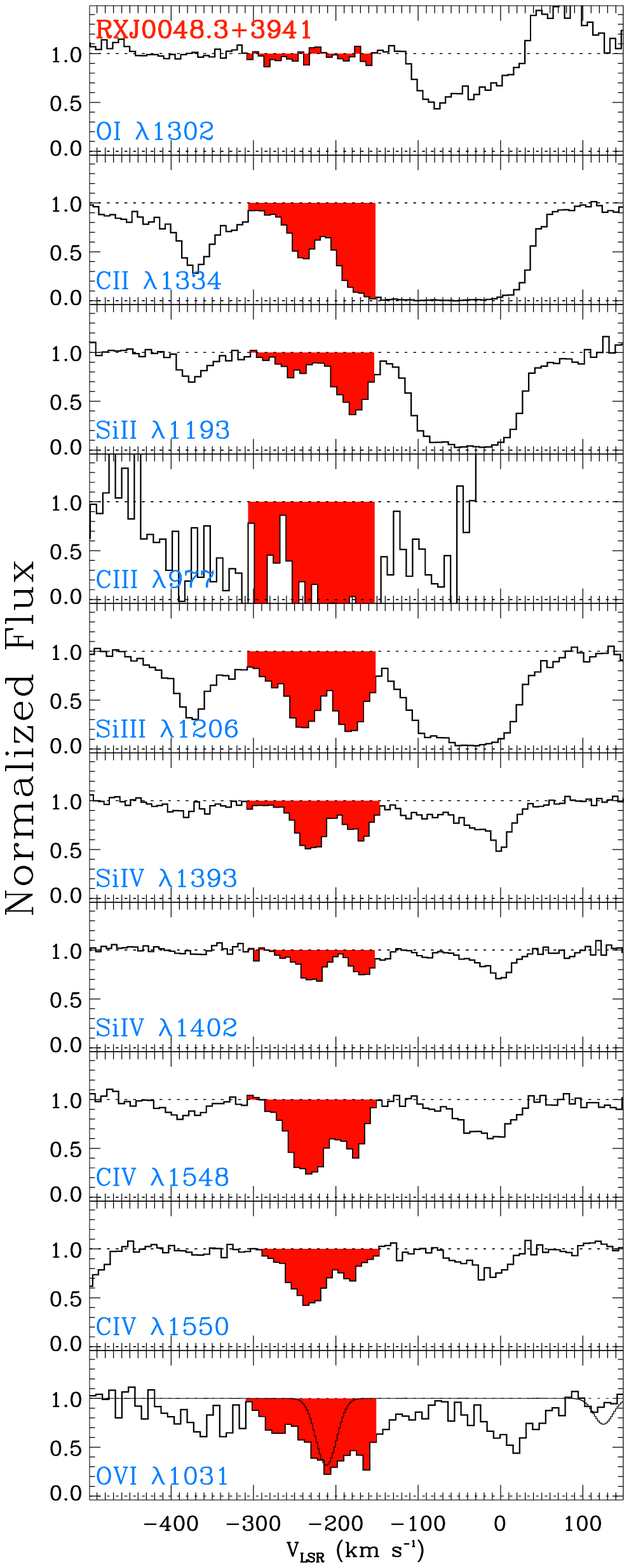}  
  \caption{Normalized profiles of  RXJ0048.1+3941 at  $R = 25$ kpc. The spectra were obtained with COS G130M and G160M, except for \ciiit\ and \ovit, which were obtained with \fuse. The red region shows the absorption from the M31 CGM gas. The MW and HVC absorption is at $v_{\rm LSR} \ga -170$ \km, while the MS absorption is at $-430 \la v_{\rm LSR} \la -310$ \km. In the \ovi\ panel, we overplot the H$_2$ model that was used to correct the \ovit\ absorption from the H$_2$ contamination. Note that \oit\ is affected by \oit\ airglow emission line at $v_{\rm LSR} \ga -100$ \km. \label{f-rxj}}
\end{figure}

\begin{figure}[tbp]
\epsscale{1.0} 
\plotone{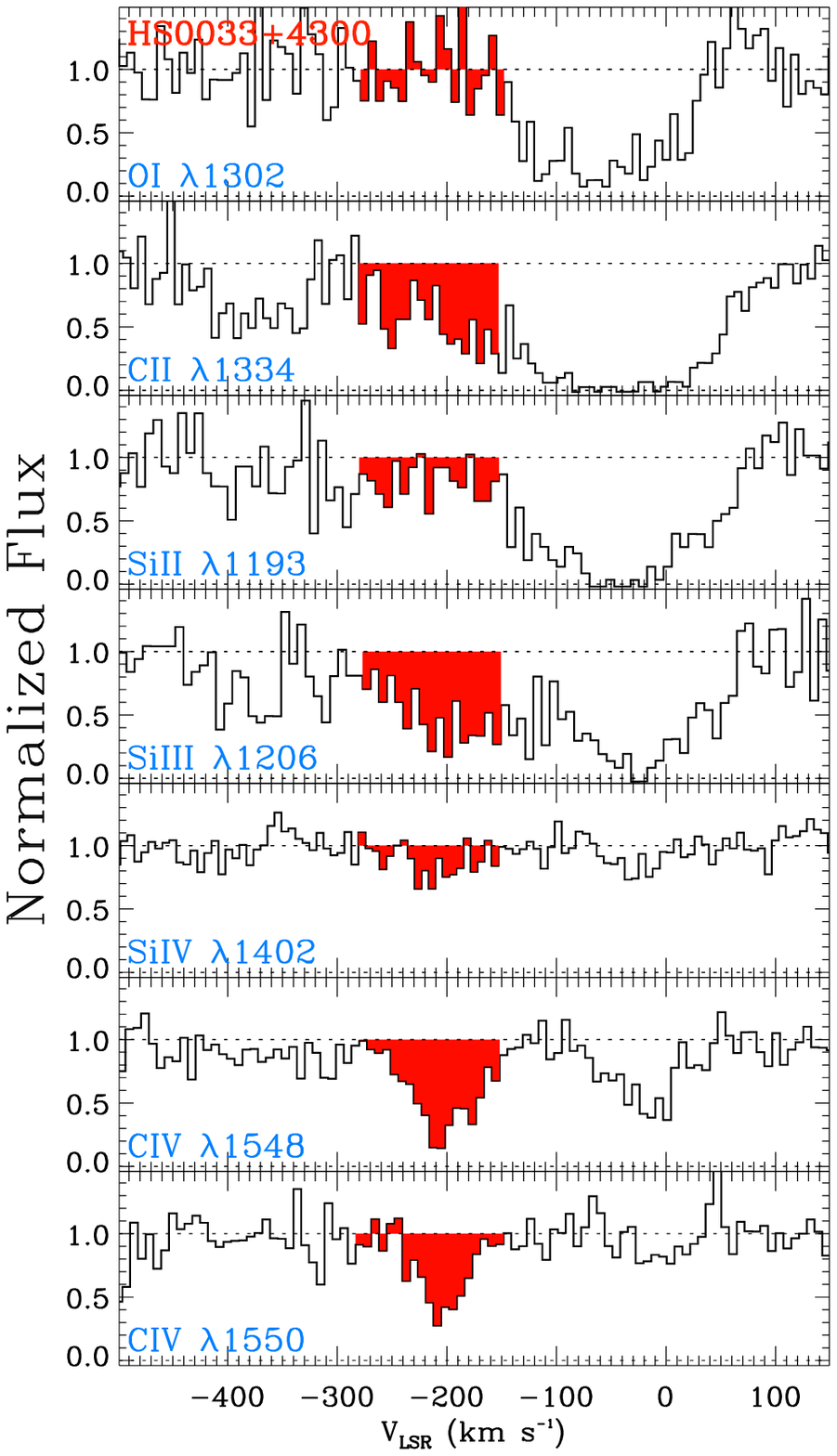}  
  \caption{Same as Fig.~\ref{f-rxj}, but for HS0033+4300 at $R = 31$ kpc. \label{f-hs0033}}
\end{figure}

\begin{figure}[tbp]
\epsscale{1.0} 
\plotone{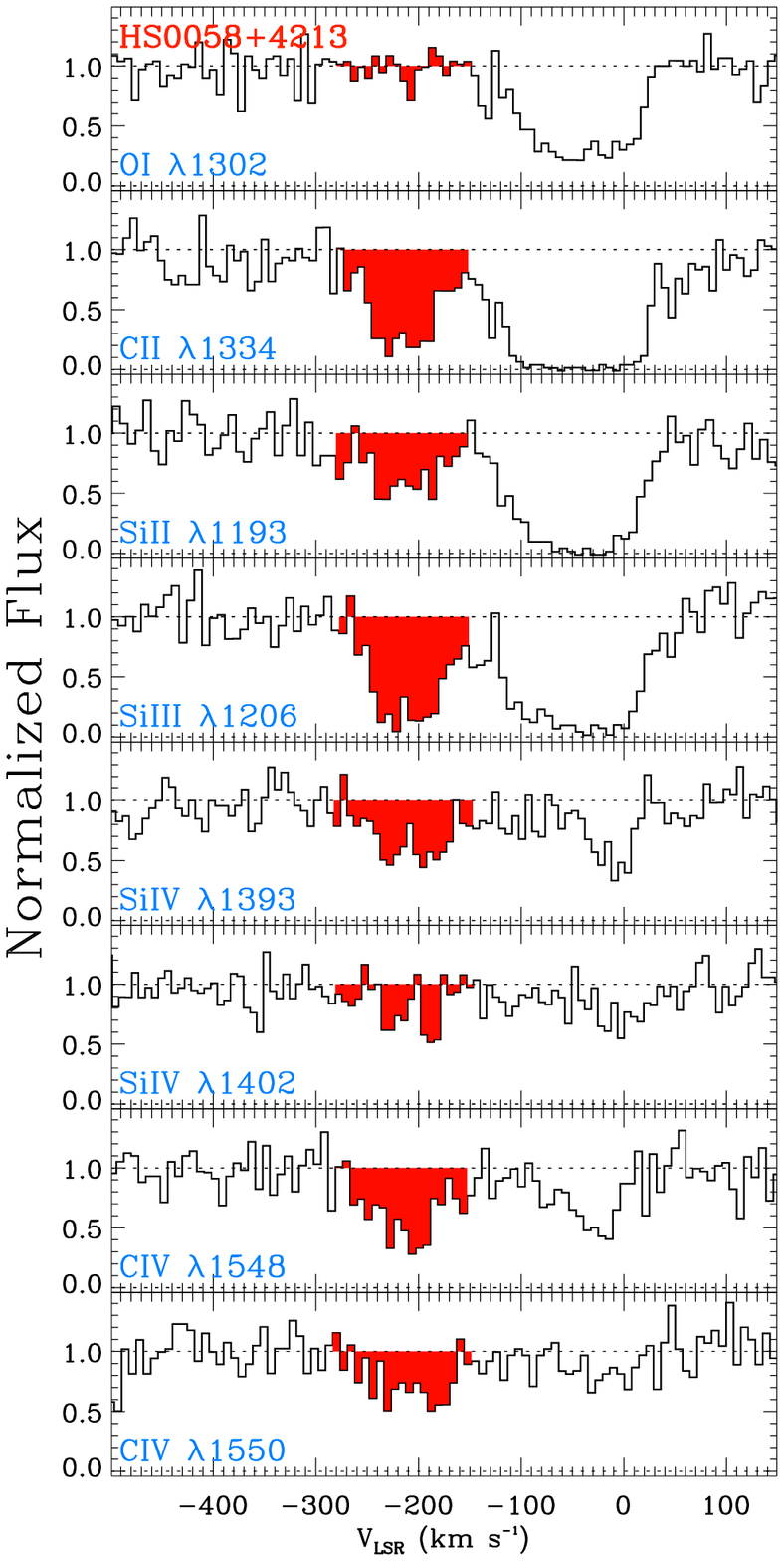}  
  \caption{Same as Fig.~\ref{f-rxj}, but for HS0058+4213 at $R =49$ kpc. There is no evidence of MS absortpion toward this sightline.  \label{f-hs0058}}
\end{figure}

\begin{figure}[tbp]
\epsscale{1.0} 
\plotone{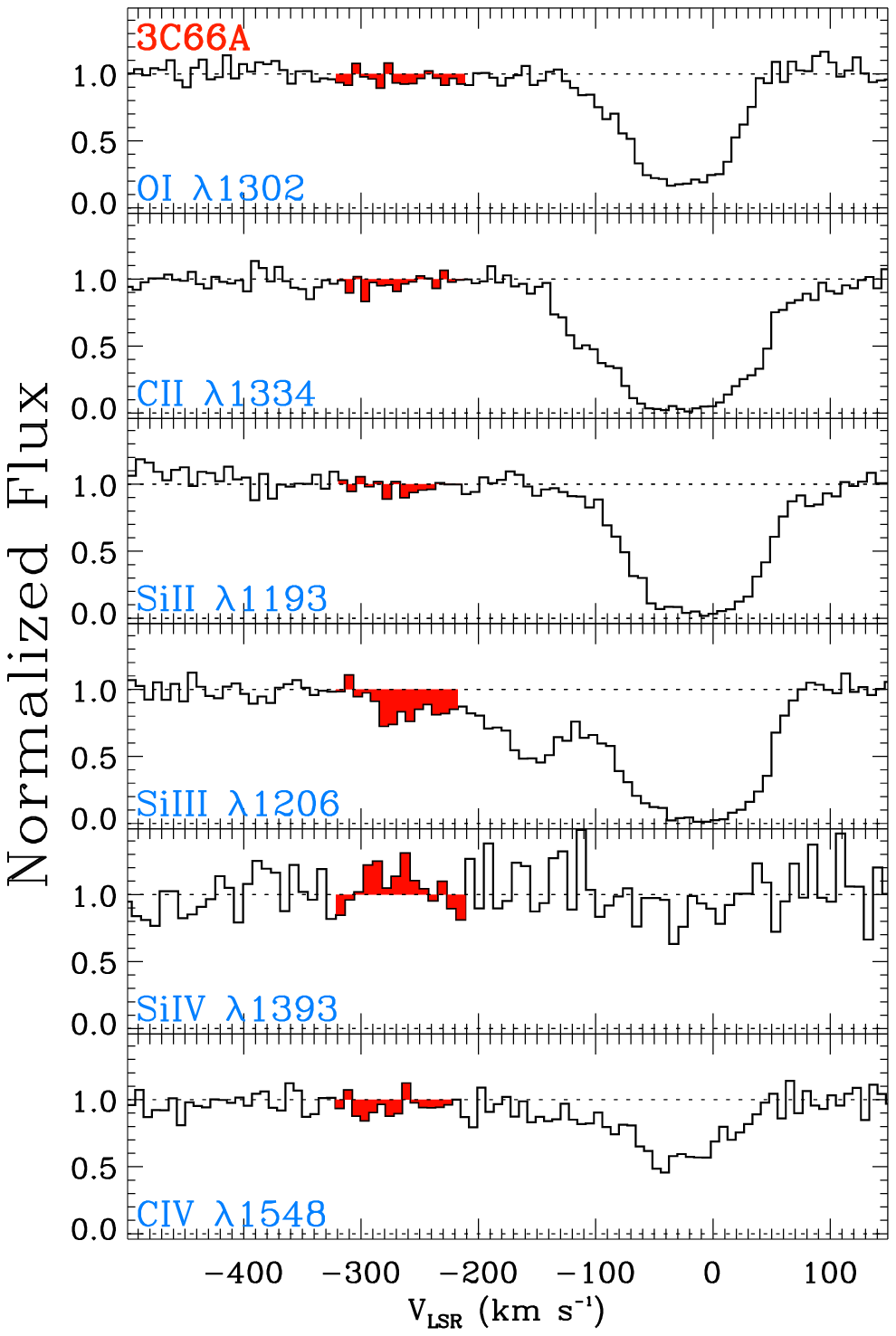}  
  \caption{Same as Fig.~\ref{f-rxj}, but for 3C66A at $R =239$ kpc. \label{f-3c66}}
\end{figure}

\begin{figure}[tbp]
\epsscale{1.0} 
\plotone{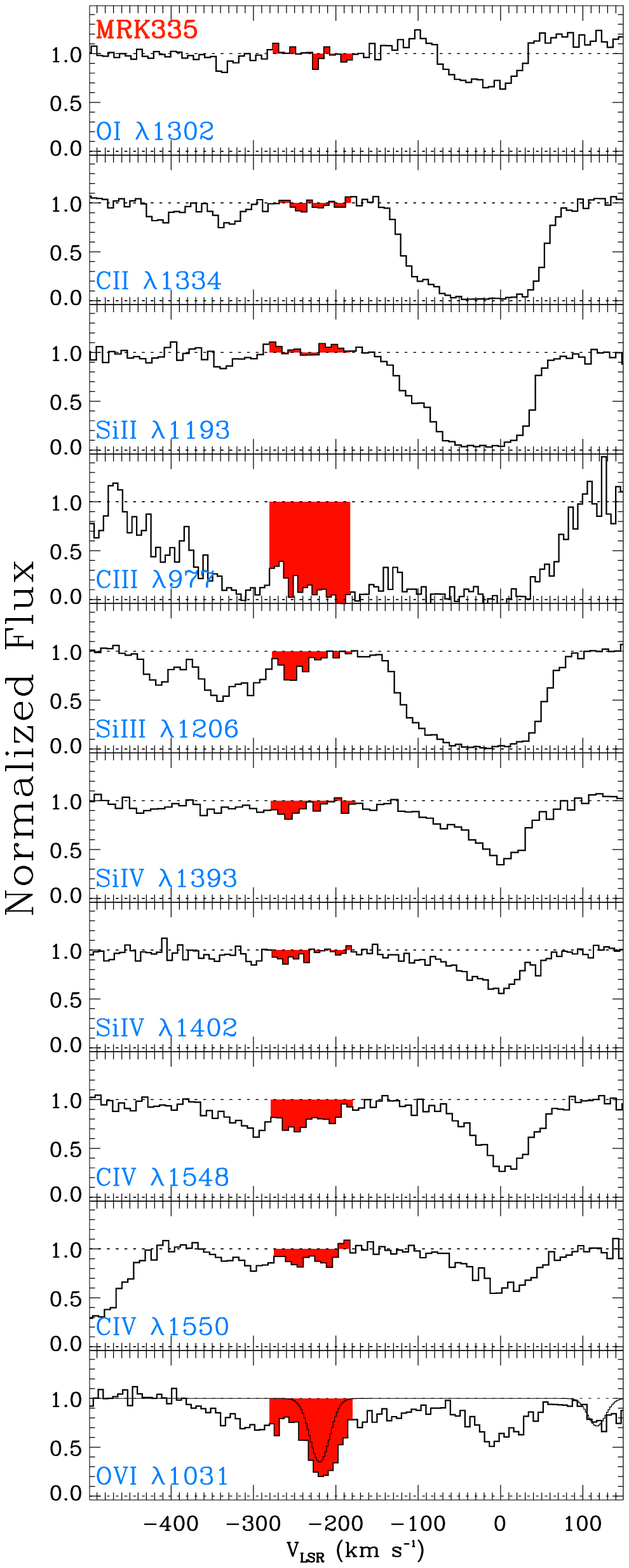}  
  \caption{Same as Fig.~\ref{f-rxj}, but for MRK335 at $R =287$ kpc.  \label{f-mrk335}}
\end{figure}

\begin{figure}[tbp]
\epsscale{1.0} 
\plotone{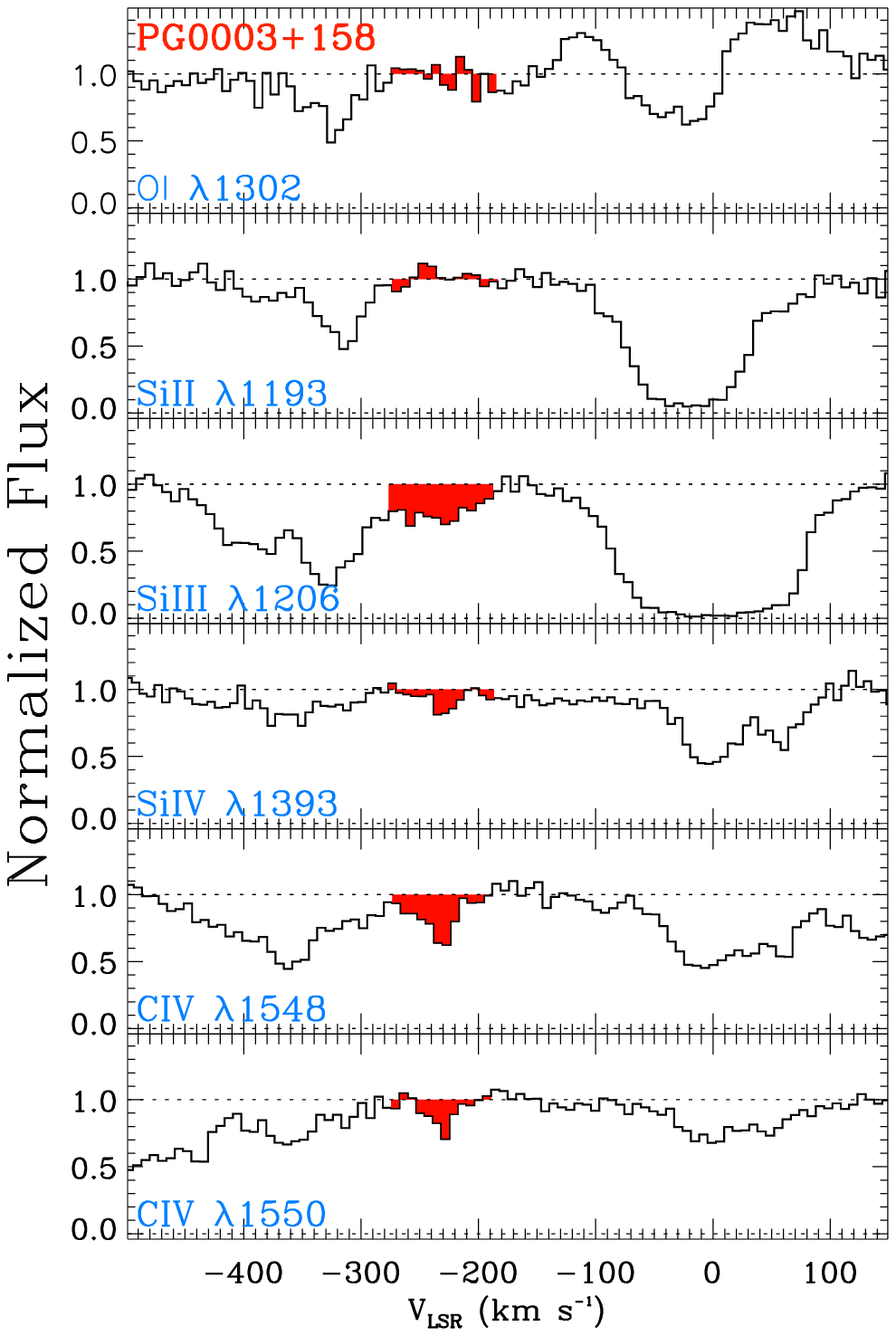}  
  \caption{Same as Fig.~\ref{f-rxj}, but for PG0003+158 at $R =334$ kpc. \label{f-pg0003} }
\end{figure}

\begin{figure}[tbp]
\epsscale{1.0} 
\plotone{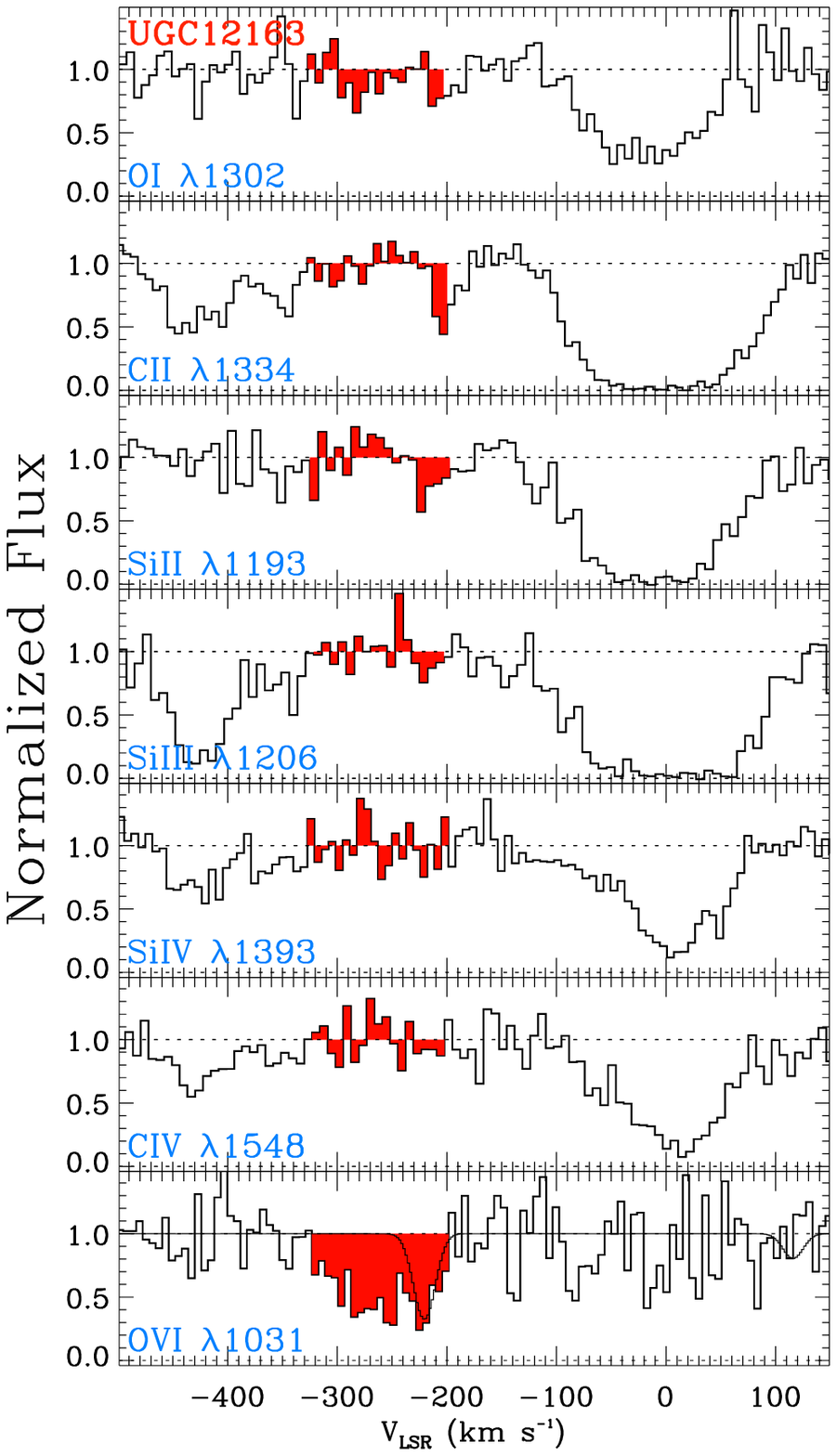}  
  \caption{Same as Fig.~\ref{f-rxj}, but for UGC12163 at $R =339$ kpc.   \label{f-UGC12163} }
\end{figure}

\begin{figure}[tbp]
\epsscale{1.0} 
\plotone{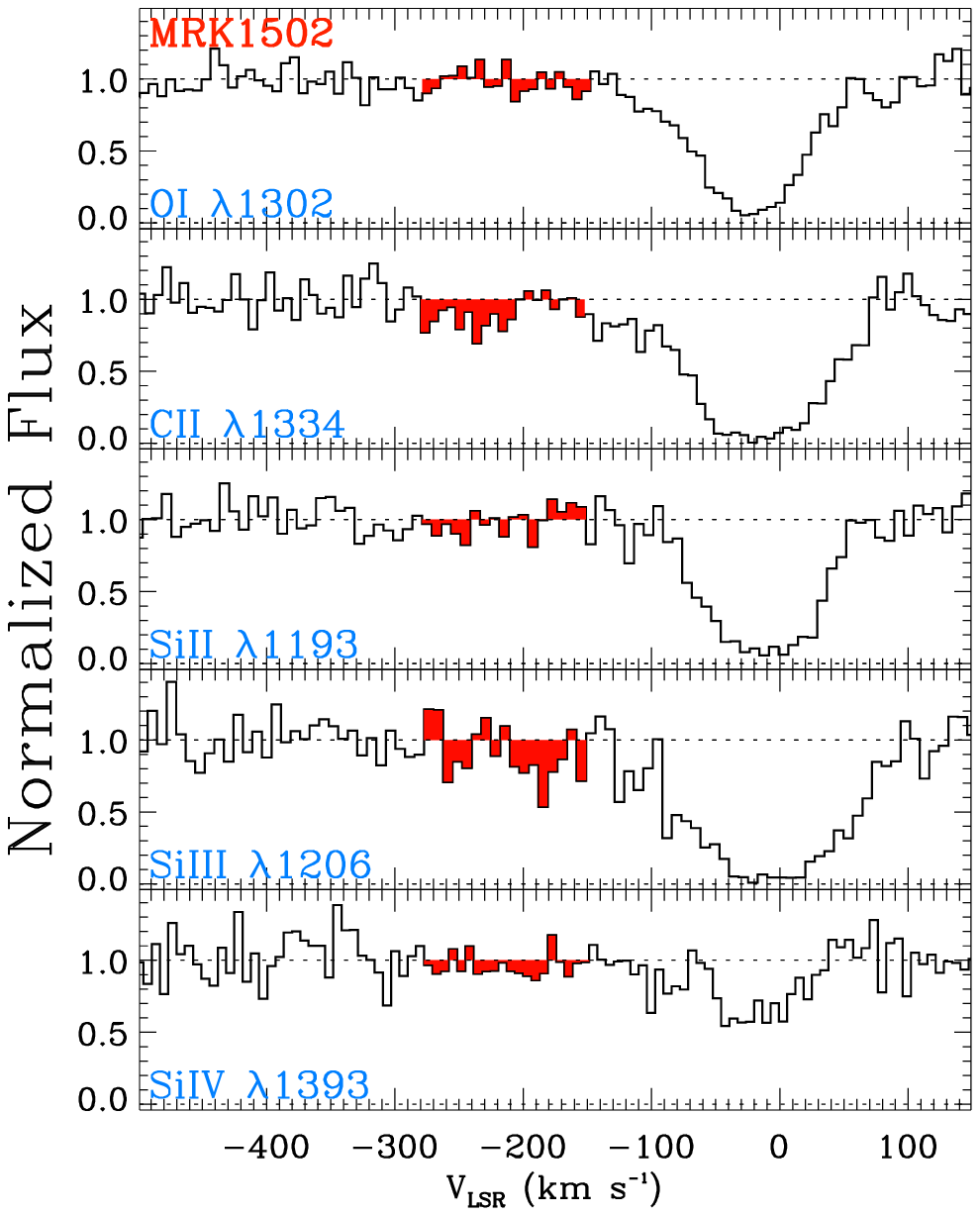}  
  \caption{Same as Fig.~\ref{f-rxj}, but for MRK1502 at $R =361$ kpc.  \label{f-MRK1502}  }
\end{figure}

\begin{figure}[tbp]
\epsscale{1.0} 
\plotone{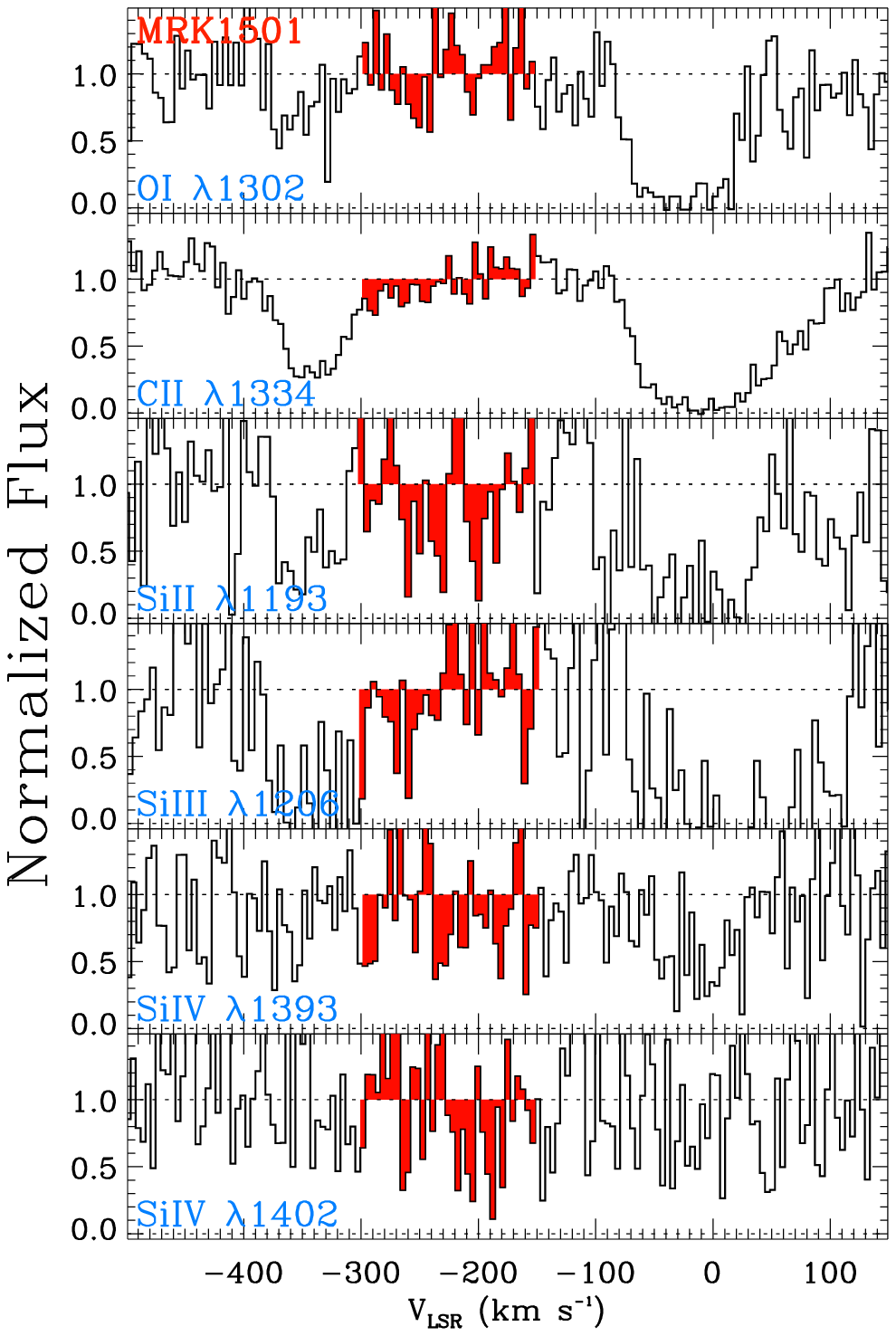}  
  \caption{Same as Fig.~\ref{f-rxj}, but for MRK1501 at $R =389$ kpc.  \label{f-MRK1501}  }
\end{figure}

\begin{figure}[tbp]
\epsscale{1.0} 
\plotone{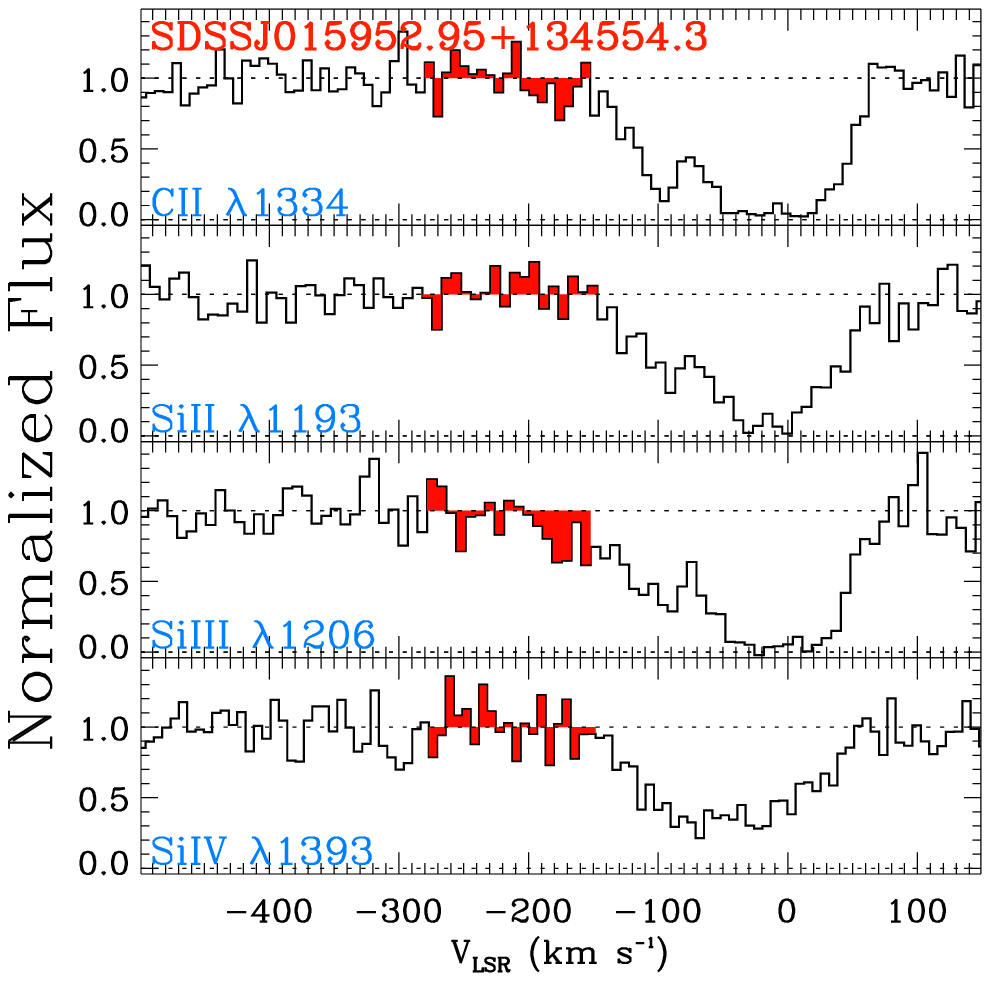}  
  \caption{Same as Fig.~\ref{f-rxj}, but for SDSSJ015952.95+134554.3 at $R =401$ kpc. \label{f-SDSSJ015952.95+134554.3}   }
\end{figure}

\begin{figure}[tbp]
\epsscale{1.0} 
\plotone{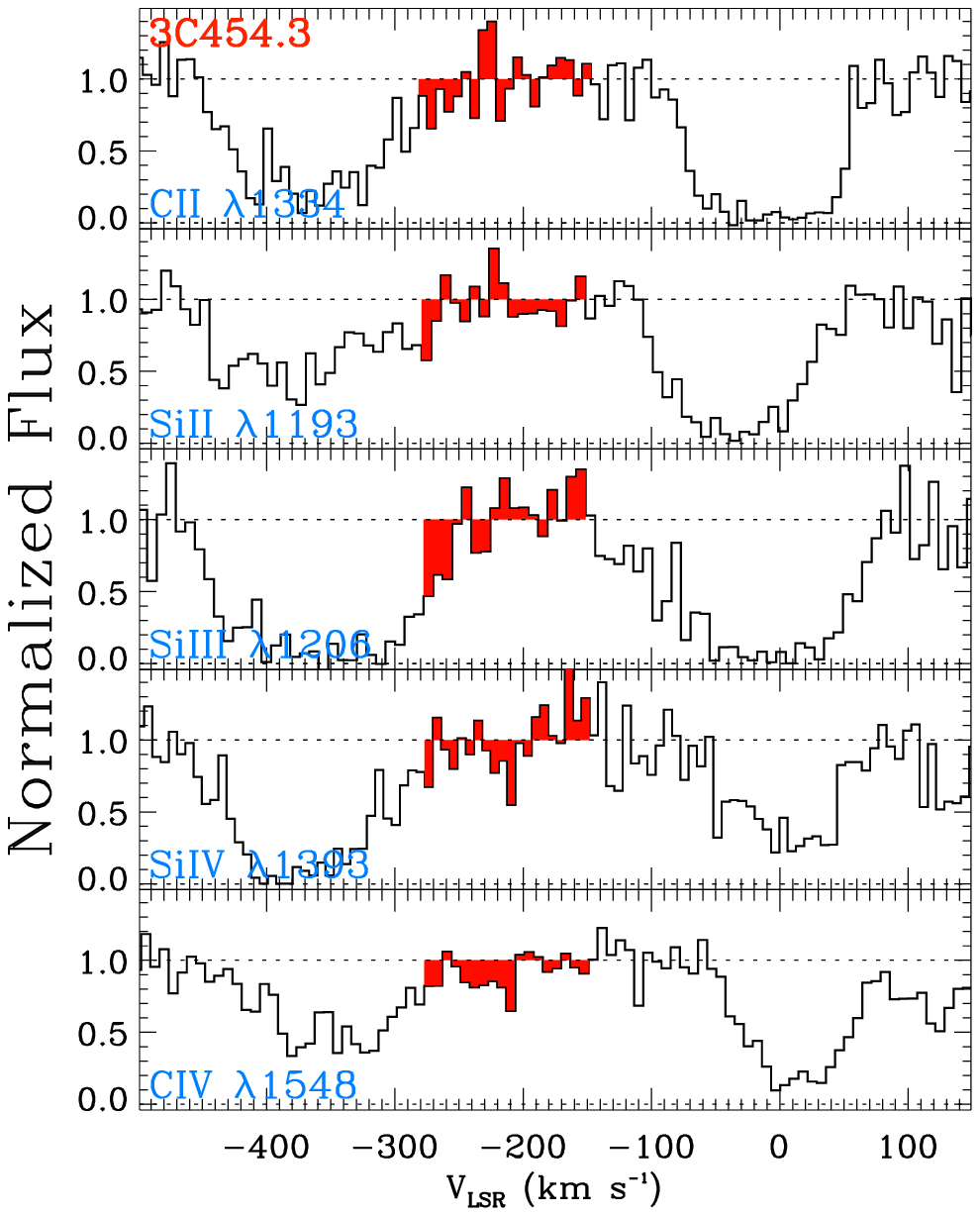}  
  \caption{Same as Fig.~\ref{f-rxj}, but for 3C454.3 at $R =424$ kpc.   \label{f-3C454.3} }
\end{figure}

\begin{figure}[tbp]
\epsscale{1.0} 
\plotone{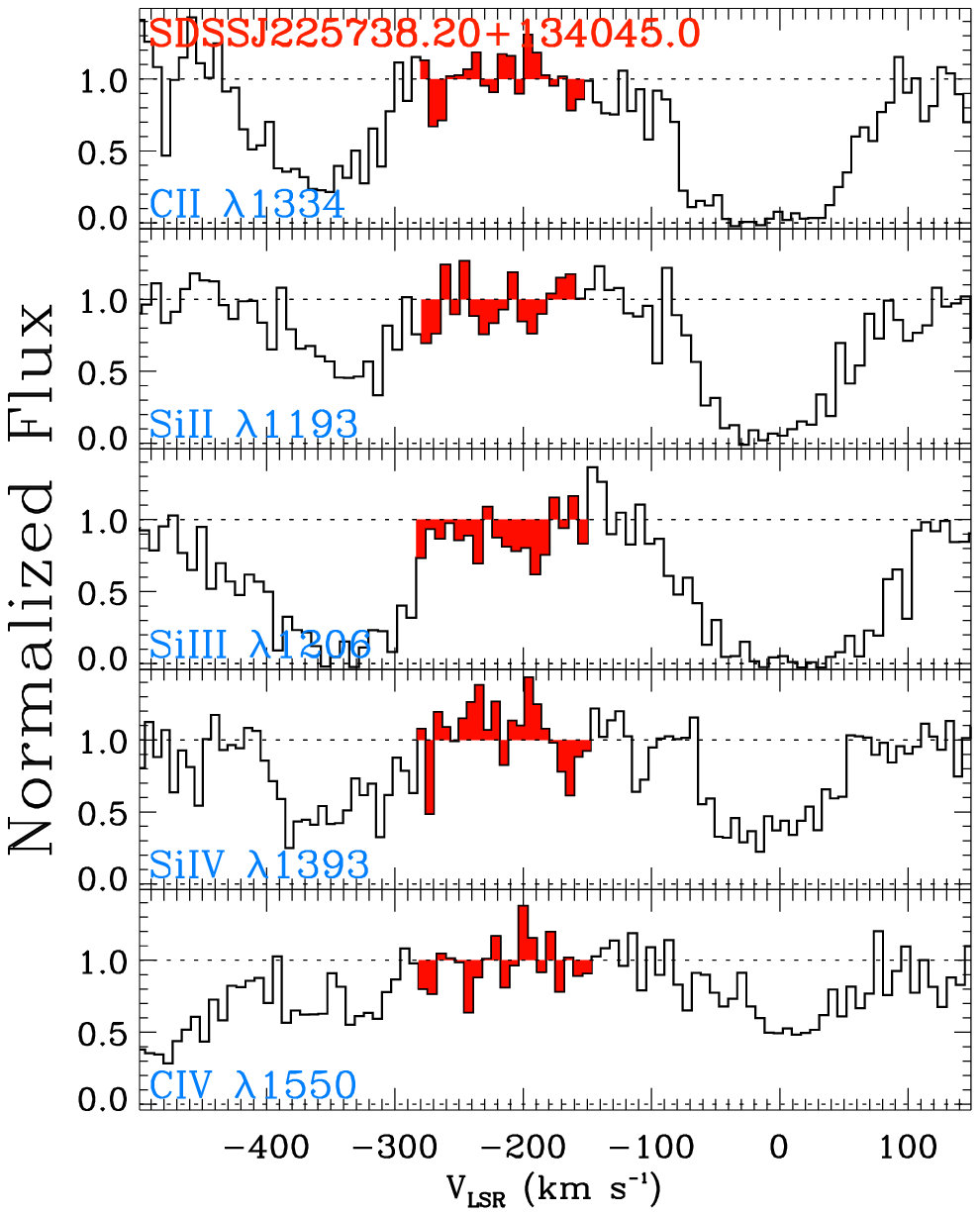}  
  \caption{Same as Fig.~\ref{f-rxj}, but for SDSSJ225738.20+134045.0 at $R =440$ kpc.  \label{f-SDSSJ225738.20+134045.0} }
\end{figure}

\begin{figure}[tbp]
\epsscale{1.0} 
\plotone{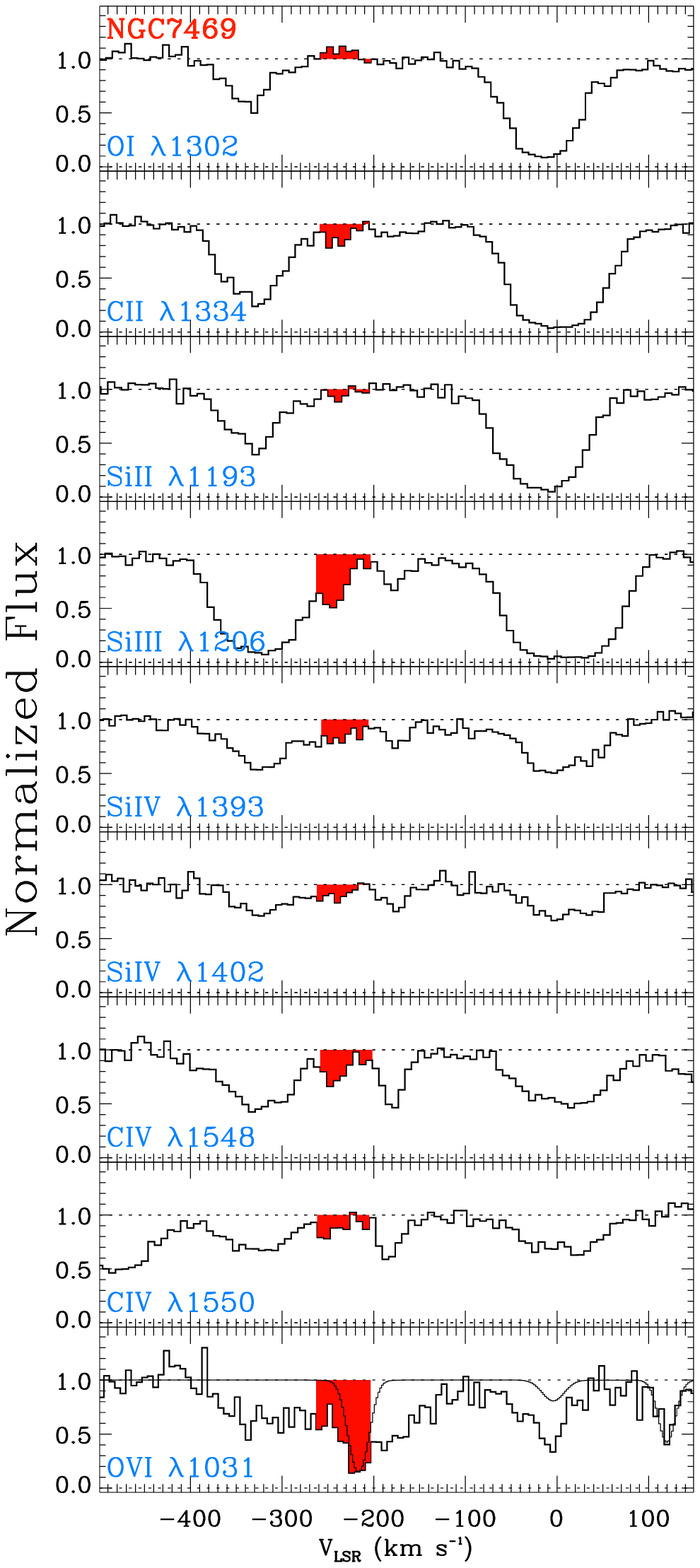}  
  \caption{Same as Fig.~\ref{f-rxj}, but for NGC7469 at $R =475$ kpc. \label{f-NGC7469}  }
\end{figure}

\begin{figure}[tbp]
\epsscale{1.0} 
\plotone{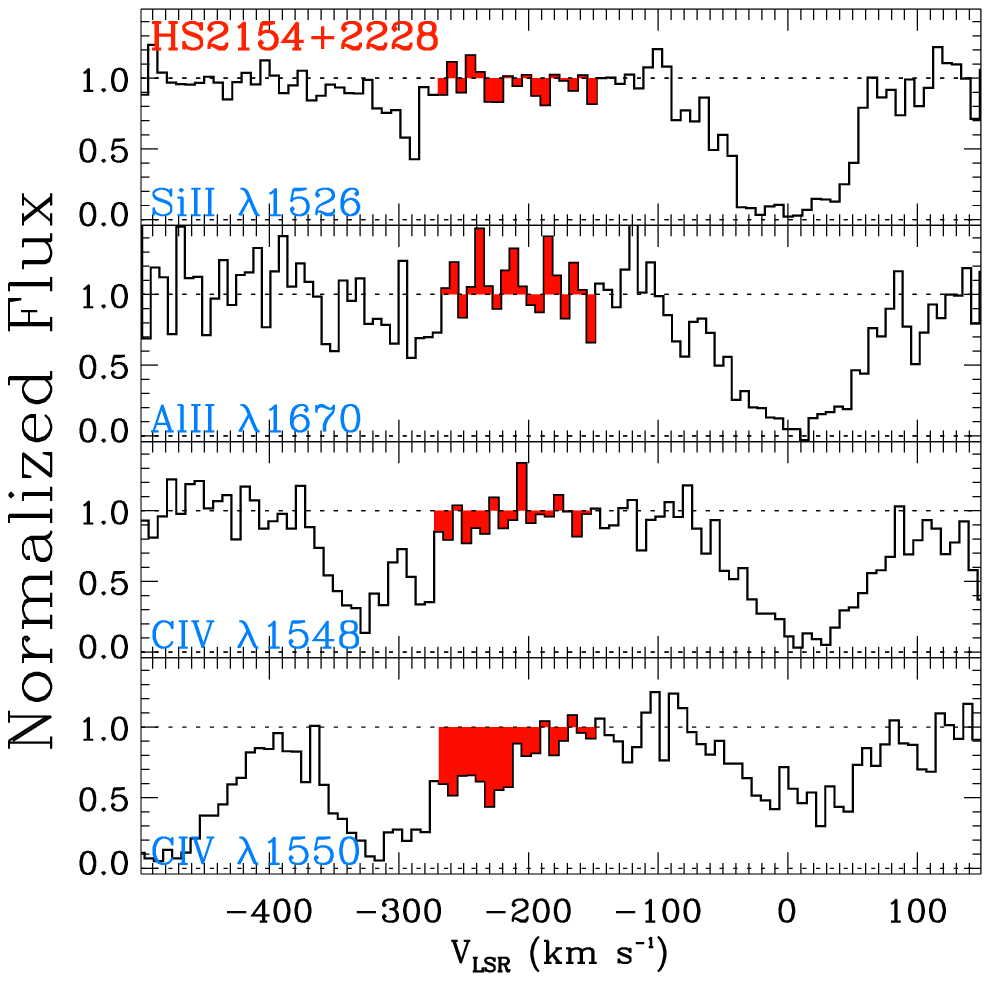}  
  \caption{Same as Fig.~\ref{f-rxj}, but for HS2154+2228 at $R =476$ kpc. \label{f-HS2154+2228}  }
\end{figure}

\begin{figure}[tbp]
\epsscale{1.0} 
\plotone{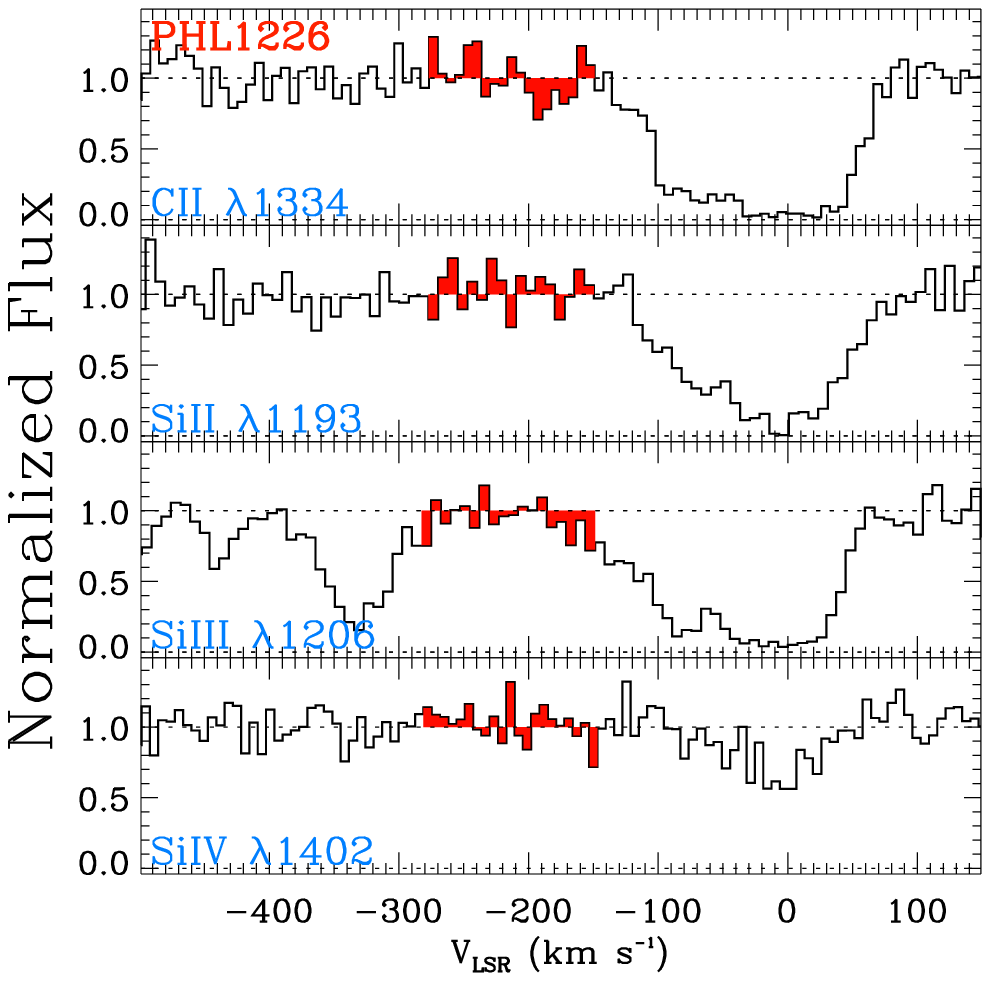}  
  \caption{Same as Fig.~\ref{f-rxj}, but for PHL1226 at $R =482$ kpc.  \label{f-PHL1226}}
\end{figure}

\begin{figure}[tbp]
\epsscale{1.0} 
\plotone{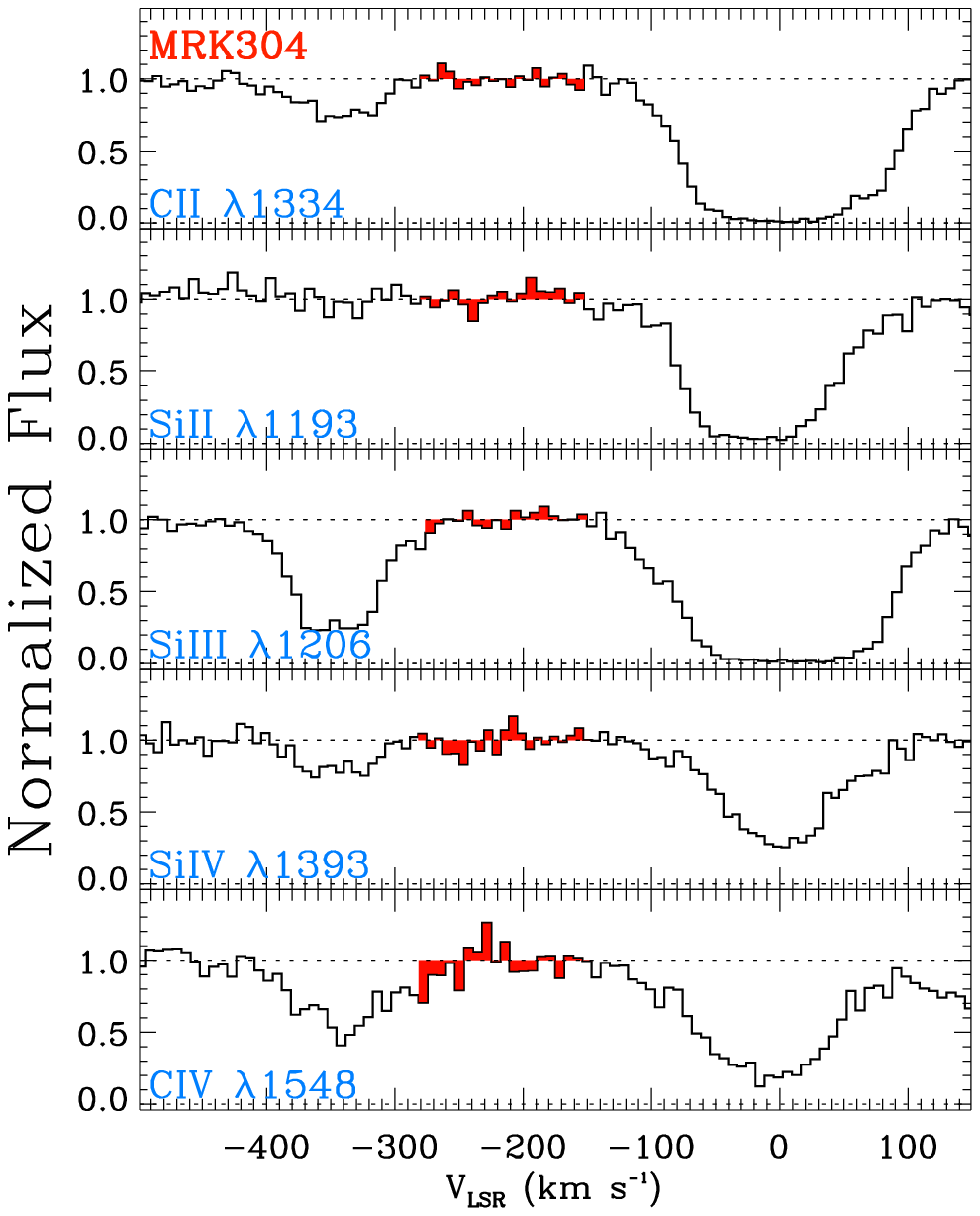}  
  \caption{Same as Fig.~\ref{f-rxj}, but for MRK304 at $R =495$ kpc.  \label{f-MRK304}}
\end{figure}

\begin{figure}[tbp]
\epsscale{1.0} 
\plotone{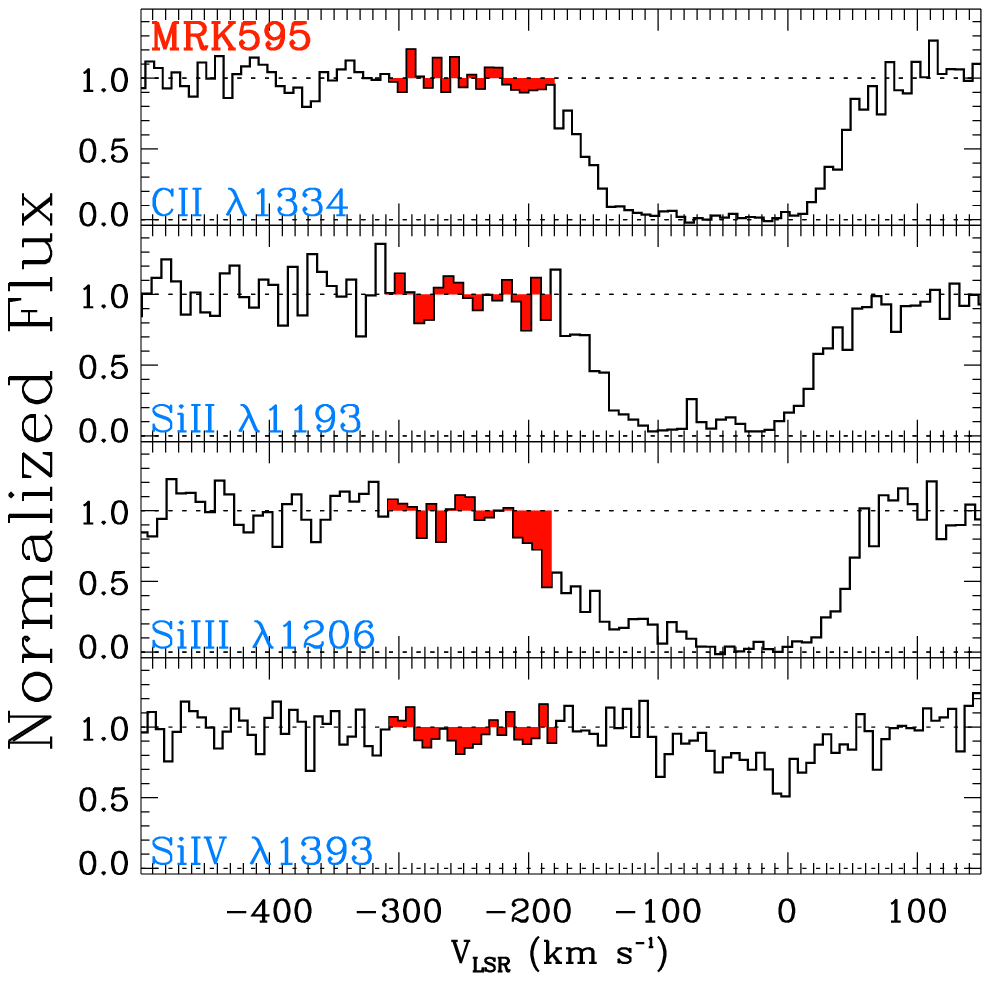}  
  \caption{Same as Fig.~\ref{f-rxj}, but for MRK595 at $R =514$ kpc.    \label{f-MRK595}}
\end{figure}

\begin{figure}[tbp]
\epsscale{1.0} 
\plotone{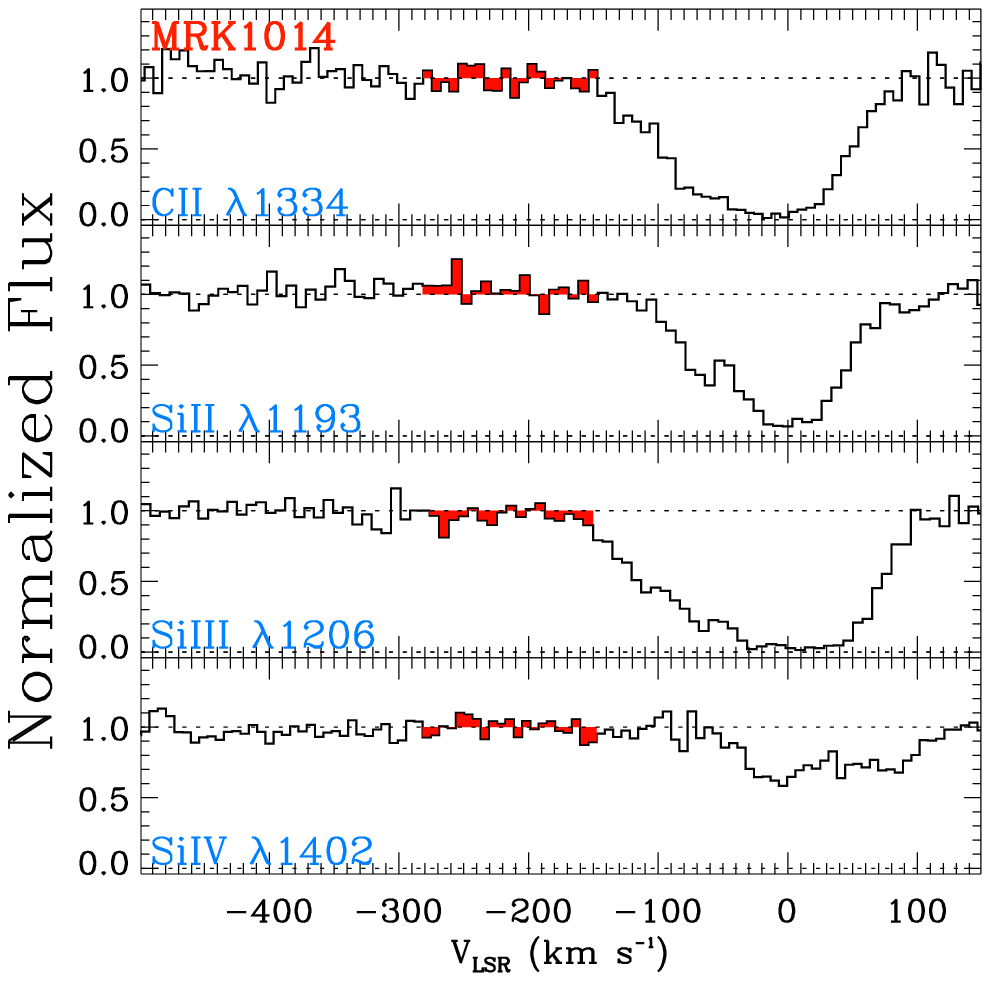}  
  \caption{Same as Fig.~\ref{f-rxj}, but for MRK1014 at $R =527$ kpc.  \label{f-MRK1014}}
\end{figure}

In this appendix, we show all  the absorption profiles used in this work as well as describe in detail each line of sight.  Figs.~\ref{f-rxj} to \ref{f-MRK1014} show the \hst/COS and \fuse\ absorption-line normalized profiles for the 18 QSO targets in the sample, sorted by increasing projected distances from the center of M31. The red region in each spectrum shows the component that is associated with the CGM of M31 (see \S\ref{s-assoc}). Absorption at $v_{\rm LSR}>-150$ \km\ is associated with the MW disk and halo. When detected, the absorption at $-430 \la v_{\rm LSR}\la -300$ \km\ is typically associated with the MS. We now give some additional information on the observed HVC components $v_{\rm LSR}\la -170$ \km\ (i.e., here defined as the gas that is not part of the Milky disk or halo) for each sightline. 

{\it - RXJ0048.1+3941}: In Fig.~\ref{f-rxj}, we show its normalized profiles. HVCs are observed at LSR velocities centered on $-380, -325$ \km\ that we associate with the MS extension and  $-242, -182$ \km\ that we associate with the M31 CGM. We estimate the total column densities of two velocity-components at $-242$ and $-182$ \km\ for the M31 CGM component. Absorption is observed for all the ions but \oi\ in the M31 CGM component. 

{\it - HS0033+4300}: In Fig.~\ref{f-hs0033}, we show its normalized profiles. HVCs are observed at LSR velocities centered on $-375$ \km\ that we associate with the MS extension and  $-204$ \km\ that we associate with the M31 CGM. In the M31 CGM component, there might be more than 1 component, but the S/N is too low to discern the exact velocity structure. Absorption is observed for all the ions but \oi\ in the M31 CGM component. 

{\it - HS0058+4213}: In Fig.~\ref{f-hs0058}, we show its normalized profiles. An HVC is observed at LSR velocity centered on $-211$ \km\ that we associate with the M31 CGM. In the M31 CGM component, there might be more than 1 component, but the S/N is too low to discern the exact velocity structure. No absorption at more negative velocity is observed. Absorption is observed for all the ions but \oi\ in the M31 CGM component. 

{\it - 3C66A}: In Fig.~\ref{f-3c66}, we show its normalized profiles. An HVC is observed at LSR velocity centered on $-256$ \km\ that we associate with the M31 CGM. Absorption is only observed in \siii\ $\lambda$1206 at nearly $7\sigma$ ($W_{\lambda} = 57.1 \pm 8.4$ m\AA). Because it is observed in only one transition, we report the column density as a lower limit to emphasize that this absorption could be contaminated by an unrelated absorber. 

{\it - PG0003+158}: In Fig.~\ref{f-pg0003}, we show its normalized profiles (\cii\ $\lambda$1334 is not shown because it is heavily contaminated by unrelated absorbers). Several HVCs are observed at $v_{\rm LSR} = -390,-325$ \km\ that we associate with the possible MS extension and at $-232$ \km\ that we associate with the M31 CGM. In the M31 CGM component, no \oi\ or singly ionized species are detected. 

{\it - UGC12163}: In Fig.~\ref{f-UGC12163}, we show its normalized profiles. Several HVCs are observed in several ions at $v_{\rm LSR} = -428,-325$ \km\ that we associate with the possible MS extension. Absorption at $-274$ \km\ is only observed in \ovi\ $\lambda$1031 that we associate with the M31 CGM. In this case there is no evidence for MW or MS \ovi\ absorption along this sightline, and hence \ovi\ is not blended as seen in other sightlines. However,  because it is observed in only one transition, we report the column density as a lower limit to emphasize that this absorption could be contaminated by an unrelated absorber. 

{\it - MRK1502}: In Fig.~\ref{f-MRK1502}, we show its normalized profiles where no HVC absorption is observed at the $3\sigma$ level.

{\it - MRK1501}: In Fig.~\ref{f-MRK1501}, we show its normalized profiles where no HVC absorption is observed at the $3\sigma$ level, but we note the S/N is very low in \siii, \siiii, and \siiv. 

{\it - SDSSJ015952.95+134554.3}: In Fig.~\ref{f-SDSSJ015952.95+134554.3}, we show its normalized profiles where no HVC absorption is observed at the $3\sigma$ level.

{\it - 3C454.3}: In Fig.~\ref{f-3C454.3}, we show its normalized profiles where no HVC absorption is observed at the $3\sigma$ level between $-300$ and $-150$ \km.  There is a strong absorption centered at $-380$ \km\ that we associate with the MS. 

{\it - SDSSJ225738.20+134045.0}: In Fig.~\ref{f-SDSSJ225738.20+134045.0}, we show its normalized profiles where no absorption is observed at the $3\sigma$ level between $-300$ and $-150$ \km.  There is a strong absorption centered at $-360$ \km\ that we associate with the MS. 

{\it - NGC7469}: In Fig.~\ref{f-NGC7469}, we show its normalized profiles. HVCs are observed at LSR velocities centered on $-330$ \km\ that we associate with the MS and  $-239$ \km\ that we associate with the M31 CGM. We note that an additional HVC is observed centered at about $-170$ \km\ and spanning the velocity interval $[-200,-110]$ \km that is likely associated with the MW halo (see \S\ref{s-assoc}). 

{\it - HS2154+2228}: In Fig.~\ref{f-HS2154+2228}, we show its normalized profiles. This target was only observed with G160M. The absorption seen in \civ\ at $-320$ \km\ cannot be confirmed because \civ\ $\lambda$1550 is contaminated. We only report a non-detection of \civ\ in the range $[-270,-150]$ \km. 

{\it - PHL1226}: In Fig.~\ref{f-PHL1226}, we show its normalized profiles where no absorption is observed at the $3\sigma$ level. The strong absorption seen in \siiii\ $\lambda$1206 at $v_{\rm LSR}< -300$ \km\ is not confirmed in any other ions and is likely not a MS component (especially since \cii\ absorption is not observed while it is typically observed in the MS component towartd other sightlines with similar \siiii\ absorption).

{\it - MRK304}: In Fig.~\ref{f-MRK304}, we show its normalized profiles where no absorption is observed at the $3\sigma$ level for the M31 CGM component. There is, however, strong absorption in \cii, \siiii, \siiv, and \civ\ at $-340$ \km\ that we associate with the MS extension. 

{\it - MRK595}: In Fig.~\ref{f-MRK595}, we show its normalized profiles where no HVC absorption is observed at the $3\sigma$ level for the MS or M31 CGM component.

{\it - MRK1014}: In Fig.~\ref{f-MRK1014}, we show its normalized profiles where no HVC absorption is observed at the $3\sigma$ level for the MS or M31 CGM component.

\clearpage

\clearpage

\clearpage

\end{appendix}

\end{document}